\documentclass{aa}
\usepackage{xcolor}
\usepackage{txfonts}
\usepackage[colorlinks, citecolor=blue]{hyperref}
\usepackage{mathtools}
\usepackage{ulem}
\usepackage{longtable} % for 'longtable' environment
\usepackage{lscape}
\usepackage{rotating}
\usepackage{multirow,multicol}
\usepackage{anyfontsize}
\usepackage{tablefootnote}
\newcommand{\bmath}[1]{\boldsymbol{#1}}
\newcommand{\mat}[1]{\bmath{\mathsf{#1}}}

\begin{document}

    \title{{\it Gaia} GraL: The GraL catalogue of gravitationally lensed quasars}
\subtitle{ X. Matched with \textit{Gaia} data, redshifts, and time delays\thanks{The full versions of Table \ref{catalogue}  (confirmed multiply imaged quasars and properties) and Table \ref{timedelay_table} (time delays) are available in electronic form at the CDS via anonymous ftp to cdsarc.u-strasbg.fr (130.79.128.5) or via http://cdsweb.u-strasbg.fr/cgi-bin/gcat?J/A+A/XX.}}
\author{
C. Ducourant\inst{1},
R. Teixeira\inst{2,1},
P. H. Vale-Cunha\inst{2,1},
L. Delchambre\inst{9},
A. Krone-Martins\inst{3,4},
J. Braine\inst{1},
L. Galluccio\inst{5,19},
J-F. Le Campion\inst{1},
O. S. Krinski-Moreira\inst{2},
S. Scarano Jr\inst{16},
C. B\oe hm\inst{6},
T. Connor\inst{6,7},
S. G. Djorgovski\inst{11},
M. J. Graham\inst{11},
P. Jalan\inst{13},
Q. Petit\inst{1},
S. A. Klioner\inst{14},
F. Mignard\inst{5},
V. Negi\inst{20} ,
J. Sebastian den Brok\inst{6},
I. Slezak\inst{5},
E. Slezak\inst{5},
C. Spindola-Duarte \inst{2},
D. Stern\inst{7},
J. Surdej\inst{9,15},
D. Sweeney\inst{6},
D. J. Walton\inst{17},
J. Wambsganss\inst{18}
}
\institute{
\inst{1} Laboratoire d'Astrophysique de Bordeaux, Univ. Bordeaux, CNRS, B18N, all\'ee Geoffroy Saint-Hilaire, F-33615 Pessac, France\\
\inst{2} Instituto de Astronomia, Geof\'isica e Ci\^encias Atmosf\'ericas, Universidade de S\~{a}o Paulo, Rua do Mat\~{a}o, 1226, Cidade Universit\'aria, 05508-          900 S\~{a}o Paulo, SP, Brazil\\
\inst{3} Donald Bren School of Information and Computer Sciences, University of California, Irvine, CA 92697, USA\\
\inst{4} CENTRA/SIM, Faculdade de Ci\'encias, Universidade de Lisboa, Ed. C8, Campo Grande, 1749-016, Lisboa, Portugal\\
\inst{5} Universit\'e C\^ote d'Azur, Observatoire de la C\^ote d'Azur, CNRS, Laboratoire Lagrange, Bd de l'Observatoire, CS 34229, F-06304 Nice cedex 4, France\\
\inst{6} Sydney Institute for Astronomy, School of Physics, The University of Sydney, NSW 2006, Australia\\
\inst{7} Center for Astrophysics Harvard \& Smithsonian, 60 Garden St., 02138 Cambridge, MA, USA\\
\inst{8} Jet Propulsion Laboratory, California Institute of Technology, 4800
Oak Grove Drive, Pasadena, CA 91109, USA\\
\inst{9} Space sciences, Technologies and Astrophysics Research (STAR) Institute, University of Li\`ege, Belgium\\
\inst{10} ARC Centre of Excellence for Gravitational Wave Discovery (OzGrav), Hawthorn, Victoria, Australia\\
\inst{11} Centre for Astrophysics and Supercomputing, Swinburne University of Technology, Hawthorn, Victoria, Australia\\
\inst{12} Division of Physics, Mathematics, and Astronomy, Caltech, Pasadena, CA 91125, USA\\
\inst{13} Center for Theoretical Physics, Polish Academy of Sciences, Warsaw, Poland\\
\inst{14} Lohrmann-Observatorium, Technische Universitaet Dresden, D-01062 Dresden, Germany\\
\inst{15} Aryabhatta Research Institute of Observational Sciences (ARIES), Manora Peak, Nainital, 263002, India\\
\inst{16} Departamento de F\'isica CCET, Universidade Federal de Sergipe,
 Rod. Marechal Rondon s/n, 49.100-000, Jardim Rosa Elze, S\~{a}o Crist\'ov\~{a}o, SE, Brazil\\
\inst{17} Centre for Astrophysics Research, University of Hertfordshire, College Lane, Hatfield, AL10 9AB, UK\\
\inst{18} Astronomisches Rechen-Institut (ARI), Zentrum fur Astronomie der Universitaet Heidelberg (ZAH), Moenchhofstr. 12-14, 69120 Heidelberg, Germany\\
\inst{19} INAF – Osservatorio Astronomico di Roma, Via Frascati 33, I-00078 Monte Porzio Catone, Italy.\\
\inst{20} Kavli Institute for Astronomy and Astrophysics, Peking University, Beijing 100871, China
}
   \date{Accepted 18/02/2026}
\abstract{Determining the Hubble constant tension requires alternative strategies, and multiply imaged quasars, with their intermediate redshifts, can potentially be used in this regard. We provide a currently complete catalogue of spectroscopically confirmed lensed quasars with ESA/{\it Gaia} astrometry and photometry, as well as redshifts and time delays when available.  In addition to the improved astrometry, the catalogue increases the number of lensed quasars by a factor of 1.5 (now 364, of which 277 are doubles and 87 are quads or triples) and significantly increases the number of lensing galaxies detected (now 218), which represents a major step forward.  Redshifts are provided for 347 quasars and 188 deflectors.  A completely new table of time delays, required for estimates of $H_0$, is presented, with 195 time delays from 73  systems.  {\it Gaia} absolute astrometry is sub-milliarcsecond and covers the entire sky. Future {\it Gaia} data releases will provide long-term photometry, which should provide many more time delays.  The catalogues as presented here enable machine-learning techniques to be trained and tested and subsequently applied to the {\it Gaia} data releases. Finally, we derive simple but homogeneous models of the 18 quadruply imaged quasars for which images of all four components are presented in {\it Gaia} DR3.}   

\keywords{Gravitational lensing: strong, Quasars: general, Astrometry, Methods: data analysis, Catalogues, Surveys}
\titlerunning{GraL -  A catalogue of gravitationally lensed quasars}
\authorrunning{C. Ducourant et al.}
\maketitle
\nolinenumbers
%***************************************************************************************************
\section{Introduction}
\label{intro}
%***************************************************************************************************

Gravitationally lensed quasars are precious tools for astrophysical and cosmological studies.
Identifying and characterising these lensed quasar systems is necessary for a broad range of applications, from time-delay cosmography to probing substructure in lensing galaxies.  
Specifically, there is now a division in the values of the Hubble-Lema\^itre constant (H$_0$) into local Universe values around 73 km/s/Mpc and cosmic-microwave-background-based values near 67 km/s/Mpc. Lensed quasars could be an important element in resolving or understanding the tension in the Hubble-Lema\^itre constant \citep{2025Colaco, 2021Freedman}.  
Until now, only a few attempts have been made to estimate H$_0$ in this way \citep{2020Wong} due to the required precision and the need for a lens model.

The {\it Gaia} satellite \citep{Prusti2016} of the European Space Agency (ESA) repeatedly covers the entire sky at high angular resolution, obtaining photometry and spectra.
In addition to the precise astrometry for which the mission was designed, {\it Gaia}
can detect images of lensed quasars even at very small separations ($\sim 250$ mas) and obtain repeated measurements of their fluxes.  

In 2018, the \textit{Gaia} GraL Gravitational Lens search team released the first version of a catalogue compiling known gravitationally lensed quasars from the literature \citep{2018Ducourant} as a reference set for both statistical studies and machine-learning validation. The ever-growing number of lensed quasars and the availability of {\it Gaia} Data Release 3 (DR3; \citealt{2023GaiaDR3, 2021GaiaeDR3, 2021GaiaeDR3corr}) and the recent Focus Product Release (FPR; \citealt{2024Krone}) motivated the construction of a new catalogue presenting the accurate \textit{Gaia} astrometry and new products related to the lenses. %

The new catalogue now contains all the spectroscopically confirmed gravitationally lensed quasars we could find in the literature. 
{\it Gaia} DR3 and FPR astrometry and photometry, as well as available redshifts from \textit{Gaia} and the Milliquas catalogue \citep{2023OJAp....6E..49F}, are included.  A literature search resulted in a list of the time delays from the image pairs.
The catalogue produced in this work gathers these data and is designed to support modelling efforts and to serve as a validation benchmark, and/or training set, for lens searches in current and forthcoming sky surveys. %}

Section~\ref{Known} presents the content of the catalogue of lensed quasars.  Section~\ref{GDR3FPR} describes the matching of the lenses with the {\it Gaia} DR3 and FPR.  Section~\ref{redshift} provides the redshifts of the lensed quasars and the deflecting galaxies. The compilation of time delays is presented in Sect. \ref{delay}, and the modelling of the quads fully measured by {\it Gaia} DR3 is presented in Sect. \ref{model}.

%***************************************************************************************************
\section{The catalogue}\label{Known}
%***************************************************************************************************

The current catalogue was compiled by combining the confirmed lenses from the first version of the catalogue (see \citealt{2018Ducourant} for the origin of the data) with all discoveries published since 2018, as extensively referenced in the footnote of Table \ref{catalogue}. We provide a comprehensive list of spectroscopically confirmed lenses together with their corresponding discovery publications. The size of the lenses, ranging from 0.2" to 59.9", is also provided. For each component of every lens, we include the \textit{Gaia} astrometry — that is, absolute coordinates rather than offsets relative to a reference. We therefore provide astrometric data from \textit{Gaia} DR3, the \textit{Gaia} FPR, and the discovery papers, and define the `best coordinates' (ra\_best, dec\_best) parameters as follows: \textit{Gaia} DR3 coordinates are used by default; if unavailable, \textit{Gaia} FPR values are adopted; and if neither is available, we used the \textit{Hubble} Space Telescope astrometry from \cite{2023MNRAS.518.1260S} if it exists. Finally, if none of these sources provides astrometry, the `best coordinates' correspond to the values reported in the discovery paper. The origin of these `best' coordinates is indicated in the `astrometry\_best' parameter.

The catalogue also provides redshifts from Milliquas \citep{2023OJAp....6E..49F}, \textit{Gaia} DR3 (QSOC redshifts; \citealt{2023Delchambre}), and the Southern-Hemisphere Spectroscopic Redshift Compilation (SHSRC; \citealt{specz_compilation_zenodo}). The discovery publications and the origin of redshifts are also indicated (see the footnote of Table \ref{catalogue} for an extensive list of references). Finally, the catalogue includes all published time delays.
%, which have never before been compiled in a single source.
These time delays have been carefully matched to the component numbering used in our dataset, saving the user the difficulty of cross-matching component identifications across multiple publications. Tables \ref{catalogue} and \ref{timedelay_table} present extracts of the catalogue. Additionally, a cross-match of the table has been performed with the AllWISE catalogue, and the (W1,W2,W3,W4) magnitudes, and errors are reported in our catalogue.

The first version of the catalogue \citep{2018Ducourant} included 233 confirmed lenses from the literature from before 2018, of which 43 were quads. Candidate lenses were also included. However, many authors have since published extensive lists of candidates, most of which are unlikely to be confirmed. As a result, we now track only spectroscopically confirmed gravitational lens systems.

\begin{figure}[!htp]
\begin{center}
\includegraphics[width=0.45\textwidth]{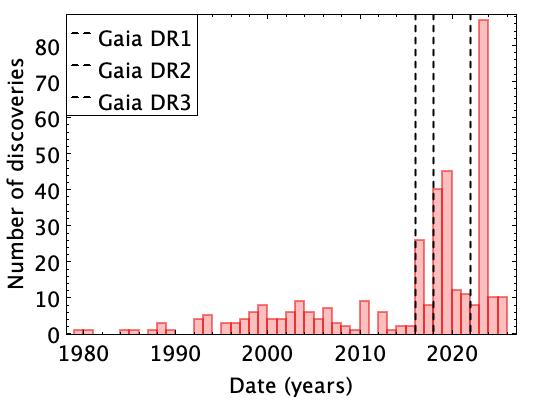}
\caption{Number of confirmed lensed quasars discovered per year. The lines correspond to the date of publication of \textit{Gaia} DR1 (September 2016), DR2 (April 2018), and DR3 (June 2022).
\label{dates_discovery}}
\end{center}
\end{figure}

 Figure~\ref{dates_discovery} shows that the number of discoveries of lensed quasars has increased rapidly over time. The vertical lines indicate the \textit{Gaia} data releases. Before \textit{Gaia} DR1, the discovery rate was approximately 5-10 per year, but immediately after its second release, it increased to 40 per year and doubled again after DR3. Among the largest contributors to these discoveries are \cite{2018MNRAS.479.5060L, 2019MNRAS.483.4242L}, \cite{2019A&A...622A.165D}, \cite{2020MNRAS.494.3491L,2023MNRAS.520.3305L}, and \cite{2021ApJ...921...42S}. This increase is not only due to the delivery of \textit{Gaia} data but also to other large surveys published in the same years, for example that from the Dark Energy Spectroscopic Instrument (DESI; \citealt{2022Duncan}). 

The current catalogue contains 1\,090 sources (quasar images and galaxies) distributed in 364 systems (80 quads, 7 triples, and 277 doublets). Figure~\ref{charts}  provides charts for all the quads and triples with the identification of the A, B, C, and D components as well as deflecting galaxies. 

\begin{figure}[!htp]
\begin{center}
\includegraphics[width=0.5\textwidth]{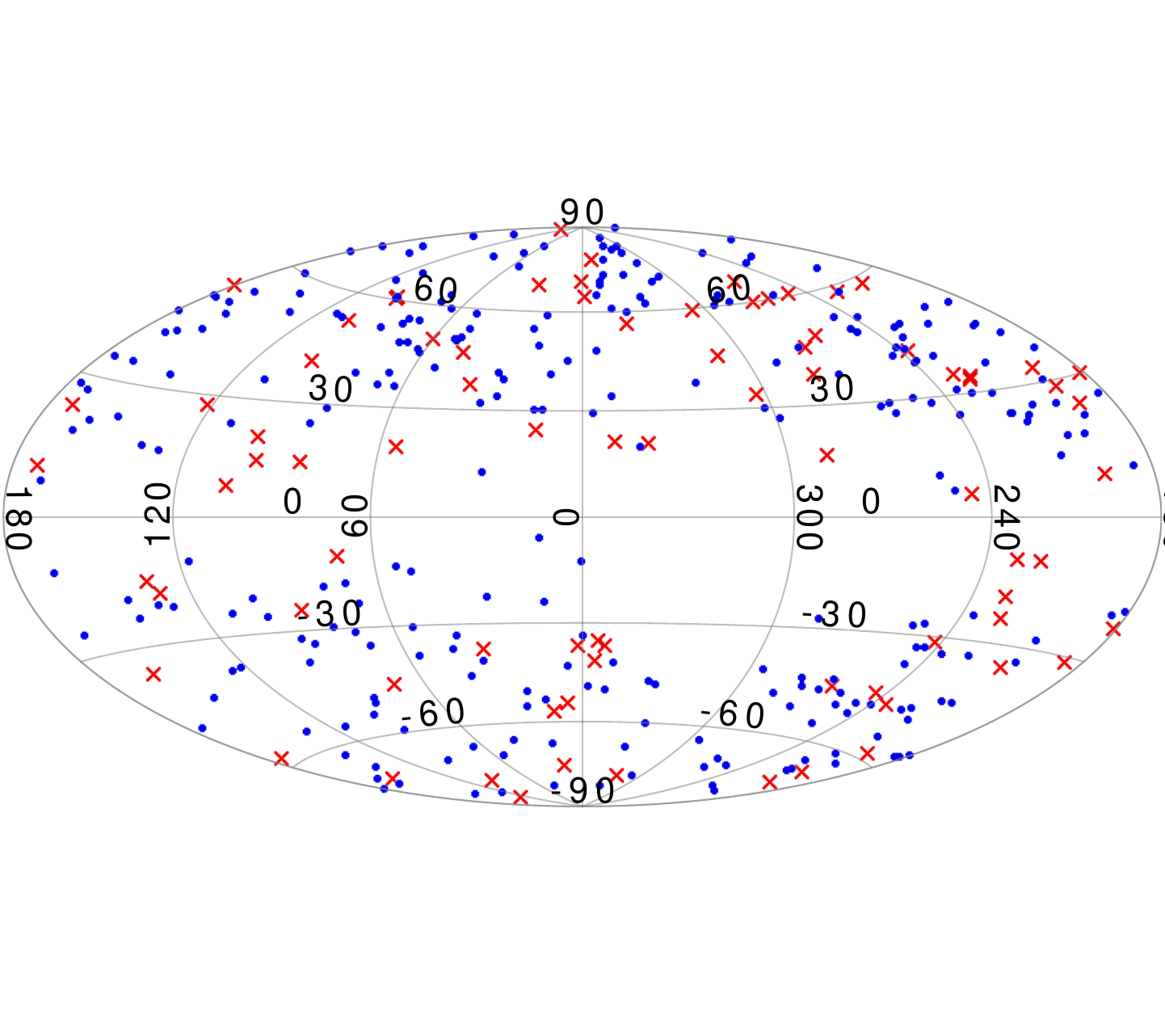}
 \caption{All-sky chart in galactic coordinates of the lensed quasars from the catalogue. Quads are indicated with red crosses and doublets with blue dots. This full-sky map uses the Hammer–Aitoff projection in galactic coordinates, with $l=b=0$ at the centre, north up, and $l$ increasing from right to left.}
\label{sky}
\end{center}
\end{figure}

The number of quads has almost doubled compared to the first release of the catalogue, and the number of doublets has increased by a factor of $\sim$1.5. An all-sky chart of the catalogue content is shown in galactic coordinates in Fig.~\ref{sky}.  

 The catalogue provides the user with the lens identifier, reference of the discovery paper, \textit{Gaia} astrometry and photometry when a match was found in DR3 or FPR, and the redshifts of the quasar and the deflecting galaxy when available, either from the discovery paper or the literature. Additionally, the catalogue has also been cross-matched with major redshift catalogues such as Milliquas, \textit{Gaia} DR3, and the SHSRC. Finally, colours from the AllWISE \citep{AllWISE} catalogue are also provided. An extract of the catalogue is shown in Table \ref{catalogue}, presenting the 18 quadruply imaged quasars fully measured by \textit{Gaia} DR3  with selected properties.

 The catalogue is available through the CDS, with access provided through Virtual Observatory-ready tools. 

%%%%%%%%%%%%%%%%%%%%%%%%%%%%%%%%%%%%%%%%%%%%%%%%
%%%%%%%%%% FIGURE DATE Discovery         %%%%%%\label{dates_discovery}
%%%%%%%%%%%%%%%%%%%%%%%%%%%%%%%%%%%%%%%%%%%%%%%%

%%%%%%%%%%%%%%%%%%%%%%%%%%%%%%%%%%%%%%%%%%%%%%%%
%%%%%%%%%% FIGURE PLOT ON SKY GL + DR3/FPR   %%%\label{sky}
%%%%%%%%%%%%%%%%%%%%%%%%%%%%%%%%%%%%%%%%%%%%%%%%

%***************************************************************************************************
\section{Match with \textit{Gaia} data}\label{GDR3FPR}
%***************************************************************************************************

\subsection{Gaia DR3 and FPR}
%***************************************************************************************************
{\it Gaia} DR3 \citep{2023GaiaDR3, 2021GaiaeDR3, 2021GaiaeDR3corr} marks a 
milestone in ESA's \textit{Gaia} mission, with the publication of tables dedicated to quasar candidates and galaxy candidates \citep{2023Coryn, 2023Ducourant}. Released in 2022, {\it Gaia} DR3 provides positions, motions, fluxes, and spectroscopic information on nearly 1.8 billion celestial objects, including 6.6 million quasar candidates. However, the catalogue remains incomplete for small angular separations \citep{2017Arenou, 2018Arenou, 2021Fabricius, 2021Torra}, which are typical for most lenses, as was the case for DR2 and was expected for early mission products. This has led some known lensed quasar images to lack \textit{Gaia} counterparts (see e.g. \citealt{2018Ducourant}). 

A \textit{Gaia} Data Processing and Analysis Consortium (DPAC) group has been working on a more complete catalogue of sources located in the immediate vicinity of a selection of 3.7 million candidate quasars. This effort resulted in the publication of the \textit{Gaia} FPR, also known as the GravLens catalogue \citep{2024Krone}. The catalogue includes the positions and magnitudes of 4.7 million sources detected within 6\arcsec\ of the candidate quasars.

 \textit{Gaia} DR3 includes the positions ($\alpha, \delta$), parallaxes ($\varpi$), proper-motions ($\mu_\alpha, \mu_\delta$), and fluxes in the $G$, $G_{\mathrm{BP}}$, and $G_{\mathrm{RP}}$ pass-bands \citep{2018Riello} along with their uncertainties. We performed a positional cross-match within a maximum angular separation of 0.5" between the astrometry found in the literature and the {\it Gaia} DR3 and FPR. From the 1\,090 individual sources (quasar images and galaxies) of our catalogue, 689 (63\% ) are measured by \textit{Gaia}. Of the 80 quads, 18 are fully detected with four images in \textit{Gaia} DR3 and 23 with measurements coming from \textit{Gaia} DR3 or FPR. 

Several quads were not completely detected despite being within {\it Gaia}'s capabilities. For example, G2237+0305 has only two of its four components published in {\it Gaia} DR3, but the two missing components are indeed present in the \textit{Gaia} FPR. The FPR astrometry and photometry are known to be less accurate than DR3 data \citep{2024Krone}, so one should consider FPR data only for sources not detected in DR3. The resolving power of the {\it Gaia} instrument is $\sim 0.2$\arcsec, but the effective angular resolution of the {\it Gaia} DR3 is poorer, $\sim 0.3$\arcsec.
The current {\it Gaia} data cross-matching, astrometric solution validation, and filtering of possible spurious sources hinder the detection of the predicted $\sim$2900 lenses \citep{2016Finet} that should be detected by the satellite. Each {\it Gaia} data release is expected to provide better effective resolution, and so merged systems in the {\it Gaia} DR3 may be separated in further data releases.

%***************************************************************************************************
\subsection{Properties}\label{astrometry}
\begin{figure}[!htp]
\begin{center}
\includegraphics[width=0.24\textwidth]{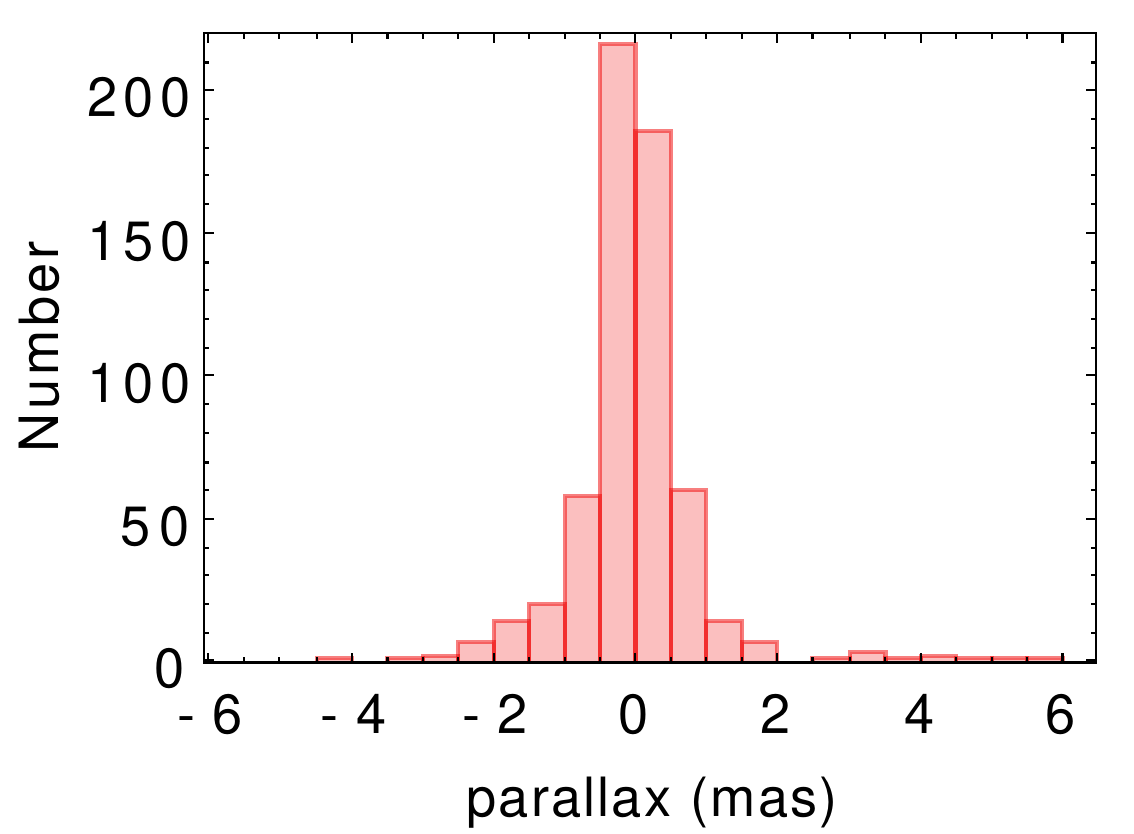}
\includegraphics[width=0.24\textwidth]{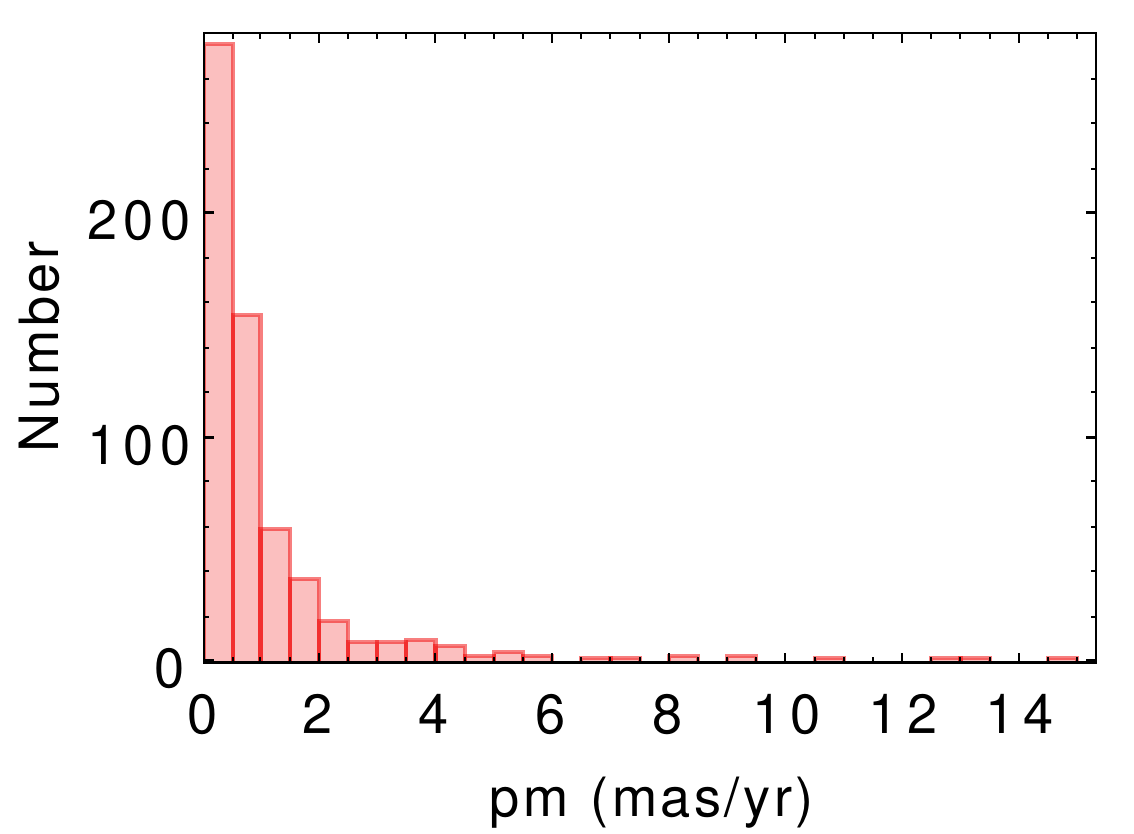}
\includegraphics[width=0.24\textwidth]{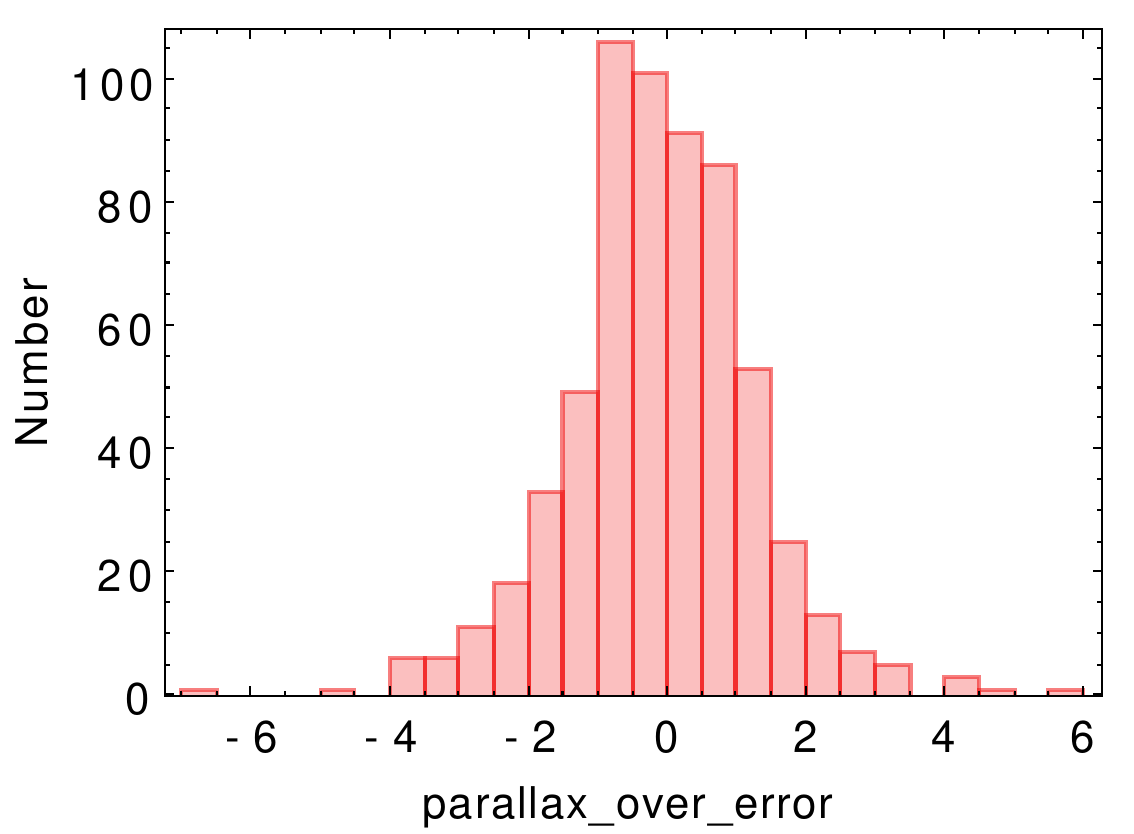}
\includegraphics[width=0.24\textwidth]{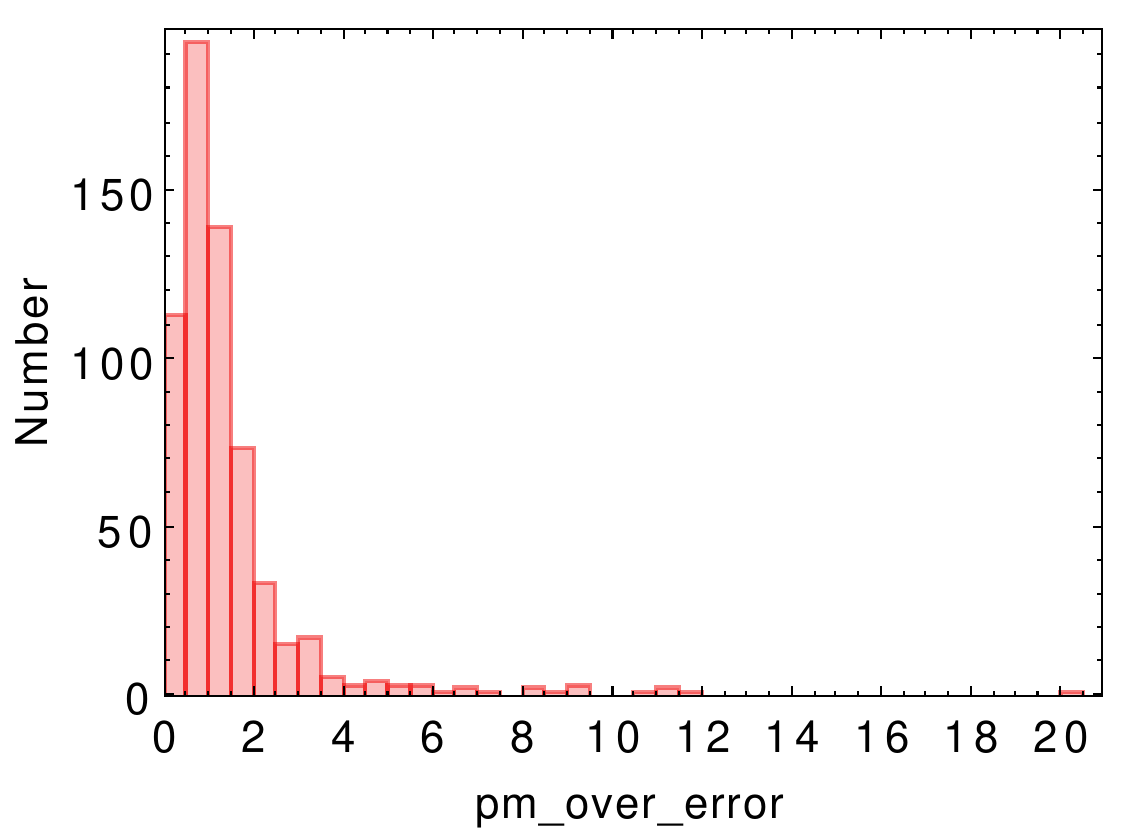}
\includegraphics[width=0.24\textwidth]{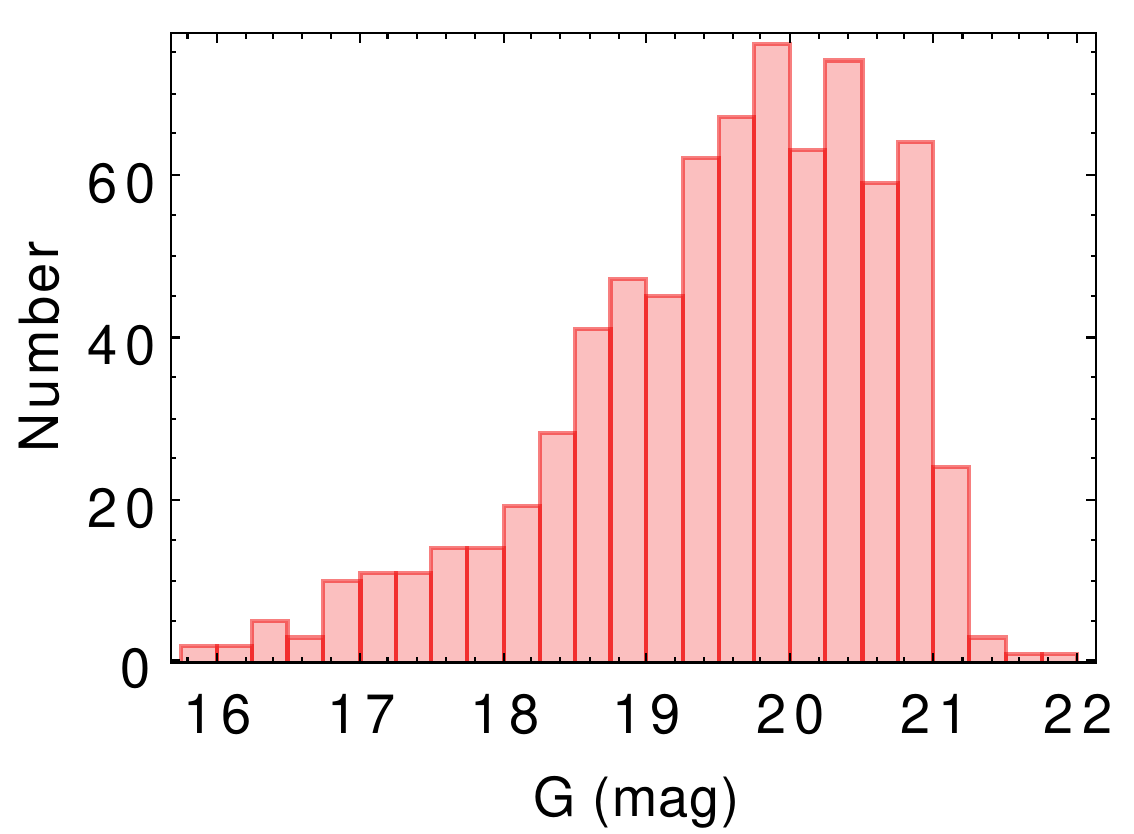}
\includegraphics[width=0.24\textwidth]{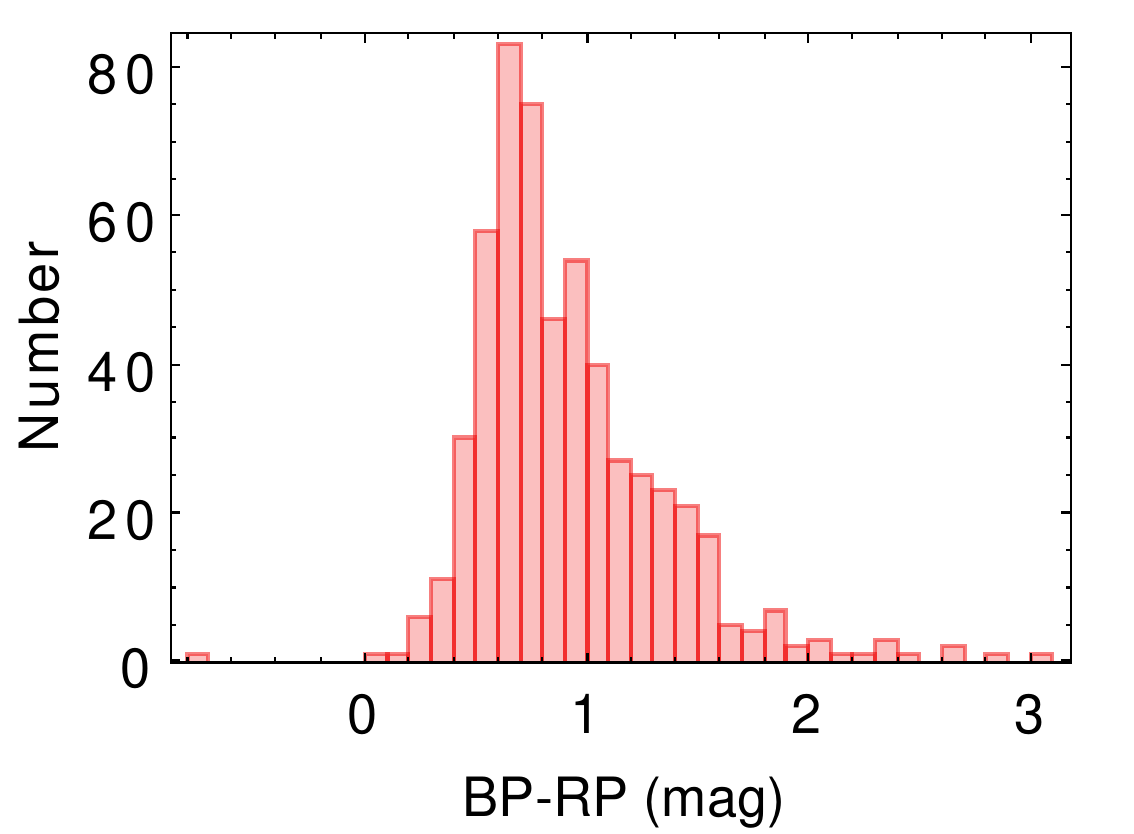}
\caption{Distributions of the astrometric and photometric parameters of sources matched with \textit{Gaia} DR3. parallax\_over\_error and pm\_over\_error stand for the parameters divided by their errors.
\label{properties}}
\end{center}
\end{figure}

%***************************************************************************************************
The statistics of the {\it Gaia} DR3 astrometric and photometric parameters are shown in Fig. \ref{properties}. As expected, the distributions of proper motions and parallaxes are centred on null values, with standard deviations smaller than 1.3 mas for parallaxes and 1.5 mas/yr for proper motions. A few large values are, however, found ($\mu_{\alpha}\cos(\delta)_{max}=-14.6 \pm2.5$ mas/yr, $\mu_{{\delta}_{max}} = -9.4 \pm 0.8$ mas/yr,  and $\varpi_{max}= -23.8 \pm3.6$ mas); this can be explained by the extended and diffuse structures of the sources, which can induce spurious parallaxes and proper motions \citep{2023Coryn, 2022ApJ...933...28M, 2022A&A...660A..16S}. They are generally near the faint limit of the {\it Gaia} detection. 

%***************************************************************************************************
Based on the known lenses, one can develop astrometric filters to search for lenses in \textit{Gaia} DR3. Catalogue sources have $\varpi - 3 \sigma_\varpi$ < 4 mas/yr and $|\,\mu\,| - 3 \sigma_{\mu}$ < 4 mas/yr, where $\mu$ stands for both $\mu_{\alpha}\cos(\delta)$ and $\mu_{\delta}$. These `soft' cuts result in some stellar contamination when applied to {\it Gaia} data, but should not result in the rejection of many genuine lens systems. Similar cuts were applied in \cite{2018A&A...616L..11K}, and also in a large, machine-learning-based, systematic blind search for lenses in {\it Gaia} DR2 \citep{2019A&A...622A.165D}.  

%%%%%%%%%%%%%%%%%%%%%%%%%%%%%%%%%%%%%%%%%%%%%%%%
%%%%%%%%%% FIGURES PROPERTIES             %%%%%%\label{properties}
%%%%%%%%%%%%%%%%%%%%%%%%%%%%%%%%%%%%%%%%%%%%%%%%

%***************************************************************************************************
\section{Redshifts}\label{redshift}
%***************************************************************************************************
%%%%%%%%%%%%%%%%%%%%%%%%%%%%%%%%%%%%%%%%%%%%%%%%
%%%%%%%%%% FIGURES REDSHIFTS        %%%%%%
%%%%%%%%%%%%%%%%%%%%%%%%%%%%%%%%%%%%%%%%%%%%%%%%
\begin{figure}[!htp]
\includegraphics[width=0.45\textwidth]{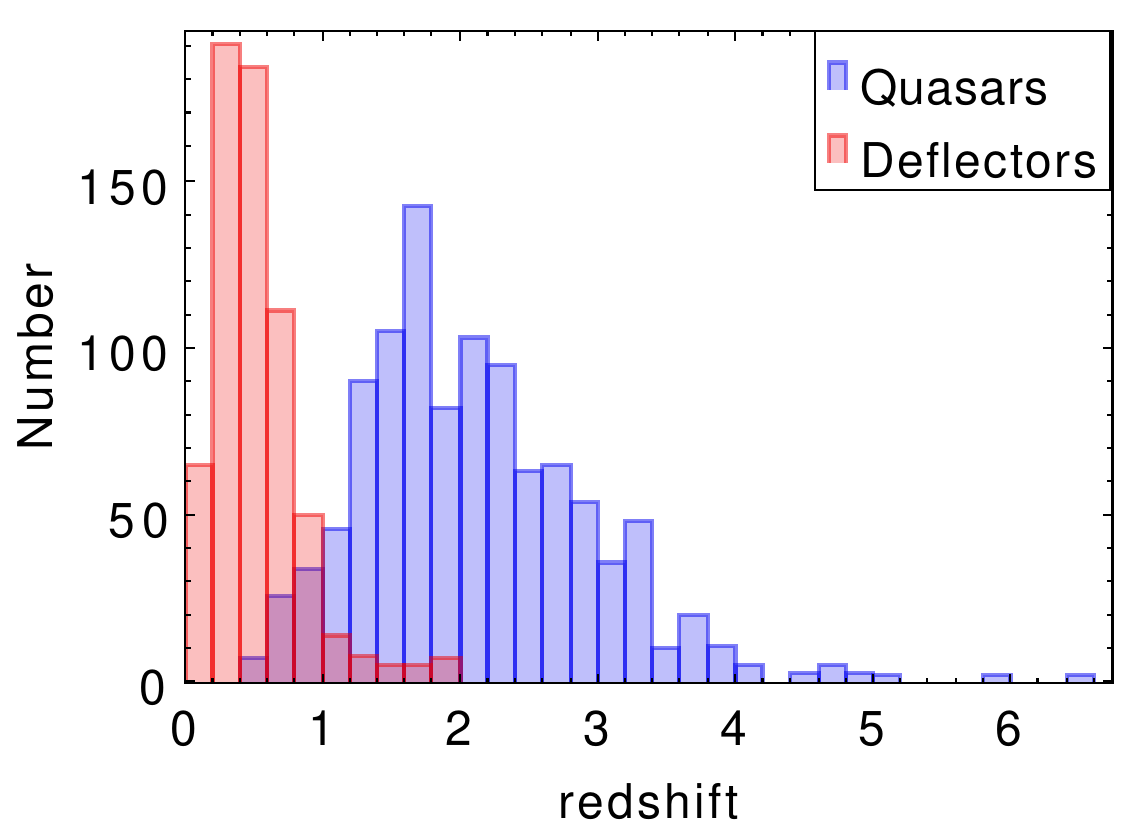}
\caption{Distribution of the spectroscopic redshifts of quasars and lensing galaxies collected from the literature (number per 0.2 redshift bins). 
\label{fig-redshift}}
\end{figure}

 We searched the literature for spectroscopic redshift information on the known gravitational lens systems (see the Table \ref{catalogue} footnote for the full list of references).
We also searched large redshift catalogues such as Milliquas \citep{2023OJAp....6E..49F}, \textit{Gaia} DR3 \citep{2023Coryn}, 
and the SHSRC \citep{specz_compilation_zenodo}. We performed a 1" positional match between the catalogue sources and these catalogues and kept the best match. 

In total, we collected 347 redshifts of lensed quasars and 189 redshifts of deflecting galaxies. Figure \ref{fig-redshift} presents the distribution of the redshifts of the quasars and the lenses. The peak of the redshift distribution of quasars is at $z_{\text{source}}\sim$1.98, and the peak of deflecting galaxies is at $z_{\text{deflector}}\sim$0.45.

%***************************************************************************************************
\section{Time delays}\label{delay}
%***************************************************************************************************

A very promising approach to resolving the tension in the Hubble-Lema\^itre constant \citep{2025Colaco, 2021Freedman} is to use time delays between quasar images to measure absolute path differences for the light.  The path length difference provides an absolute reference and hence a value for $H_0$, the Hubble-Lema\^itre constant. The uncertainty in $H_0$ arises from the uncertainty in the delay and in the model of the lens.
Determinations of $H_0$ range between the local Universe values of $\sim 73$ km/s/Mpc and the cosmology-based Wilkinson Microwave Anisotropy Probe (WMAP) or Planck values of $\sim 67$ km/s/Mpc.  Time delays are measured when there is a characteristic flux variation that can be identified in two or more light curves, meaning that quasars with very weak flux variations will not have measurable time delays.

Long time delays are generally more useful for the determination of $H_0$ because it is the fractional uncertainty that is relevant.  With its $\sim2000$ days time series, {\it Gaia} should enormously increase the number of time delays and, in particular, long time delays in its forthcoming DR4.  {\it Gaia} should complement ground-based observations, which have higher observing frequencies, generally shorter durations, and poorer angular resolution.

\begin{figure}[!htp]
\includegraphics[width=0.45\textwidth]{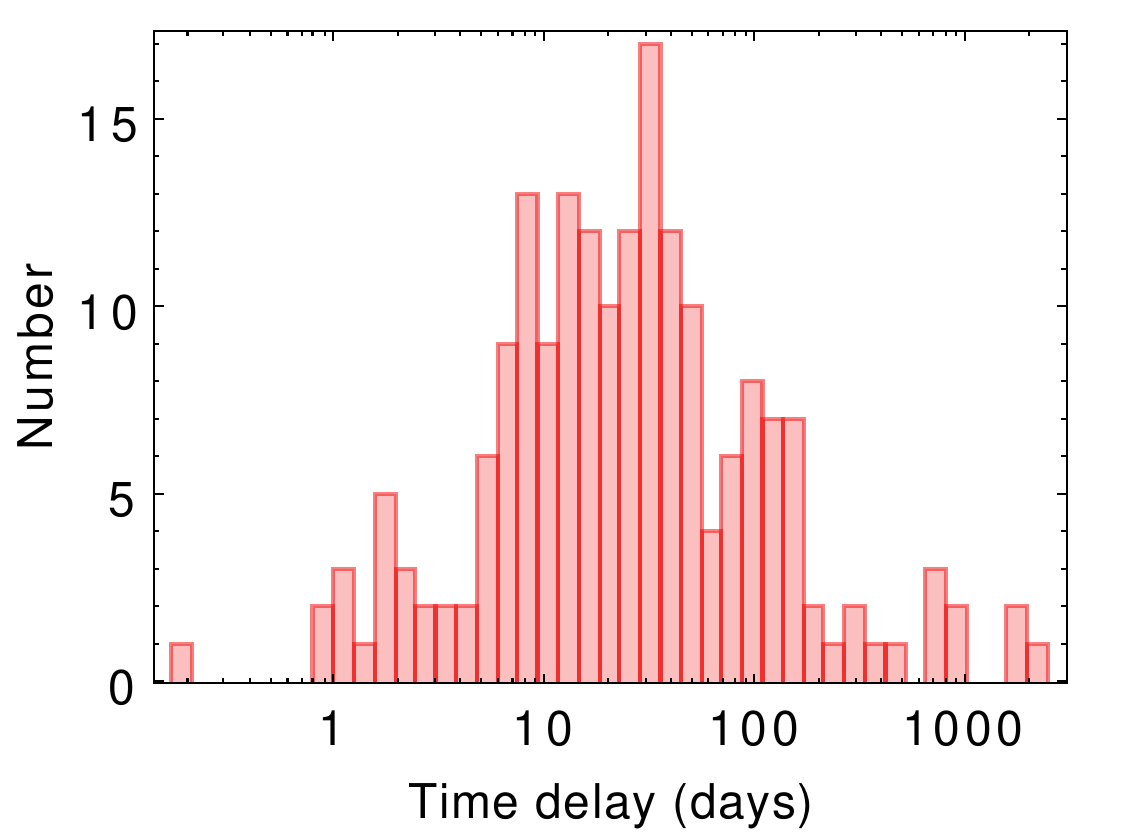}
\caption{Distributions of the values of time delays ($|\Delta$$t_{AB}|$, $|\Delta$$t_{AC}|$, ...) for the known lenses listed in Table \ref{timedelay_table}.}%
\label{timedelay}
\end{figure}

We searched the literature for published time delays of lensed quasars (see Table \ref{timedelay_table} for the full list of references) and found 73  quasars with at least one measured time delay and a total of 195 individual delays, including within a new, unpublished quad, DESJ0029--3814 (Schechter in prep.). These data are provided in Table \ref{timedelay_table}. Figure \ref{timedelay} shows the distribution of the time delays collected.

 One can see that the peak of the distribution is around 20 days, and the largest values  are around 2000 days. The number of time delays has increased rapidly recently, mostly thanks to the COSMOGRAIL and TDCOSMO publications \citep{2020Millon_a, 2020Millon_b, 2025Dux}.

%%%%%%%%%%%%%%%%%%%%%%%%%%%%%%%%%%%%%%%%%%%%%%%%
%%%%%%%%%% FIGURES TIME DELAYS        %%%%%%
%%%%%%%%%%%%%%%%%%%%%%%%%%%%%%%%%%%%%%%%%%%%%%%%

%***************************************************************************************************
\section{Modelling of the quads}\label{model}
%***************************************************************************************************

Inferring $H_0$ from time delays requires that the distribution of the mass along the line of sight be known \citep{2022A&ARv..30....8T}. However, we can usually assume that the bending of the light occurs in a plane perpendicular to the observer-lensing galaxy direction, which contains all of its projected mass. This thin lens approximation provides us with simple models for the mass distribution projected onto the lens plane (see \citealt{Keeton2001} for the most commonly used mass models). For accuracy and homogeneity, we decided to model the 18 known quadruply imaged quasars for which all four images are available in {\it Gaia} DR3. We used as input the observed positions $(x_i, y_i)$ and fluxes ($f_i$) of the lensed images $i \in \lbrace 1, \dots, 4\rbrace$ that are reproduced using a singular isothermal ellipsoid (SIE) lens model \citep{Kormann_SIE_1994} in the presence of an external shear \citep{1987ApJ...312...22K}, hereafter the SIE+shear lens model. The predicted image positions $(\hat{x}_i, \hat{y}_i)$ and amplifications ($\hat{\mu}_i$) associated with a given position of the source $(x_s, y_s)$ were computed using the {\tt lenstronomy} software \citep{2018PDU....22..189B}. We summarise here the parameters of the SIE+shear lens model that are specific to the {\tt lenstronomy} software; a more detailed description is presented in \cite{2021JOSS....6.3283B}. The SIE+shear lens model is defined through a dimensionless surface mass density, 
\begin{equation}
    \kappa(x, y) = \frac{\theta_{E}}{2} \left(q \, \left(x-x_g\right)^2 + \frac{\left(y-y_g\right)^2}{q} \right)^{-\frac{1}{2}}
    \label{eq:model_sie_shear_kappa}
,\end{equation}
where $\theta_E$ is the circularised Einstein radius; $q$ is the ratio of the minor to the major axis; $(x_g, y_g)$ is the position of the lensing galaxy; and $(x,y)$ is a coordinate system respectively aligned with the minor and major axis of the lens. The system orientation ($\phi$), shear orientation ($\phi_\gamma$), and strength ($\gamma$) are related to the complex ellipticity moduli $(e_x, e_y)$ and shear components $(\gamma_x, \gamma_y)$ presented in Table \ref{tbl:lens_modelling_parameters} through
\begin{equation}
    \begin{pmatrix}e_x \\ e_y\end{pmatrix} = \frac{1 - q}{1 + q} \;\begin{pmatrix}\,\cos(2 \phi) \, \\ \, \sin(2 \phi) \,\end{pmatrix} \hspace{0.5cm}\text{and}\hspace{0.5cm}
    \begin{pmatrix}
        \gamma_x \\ \gamma_y
    \end{pmatrix} = \gamma \; \begin{pmatrix}
        \, \cos(2 \phi_\gamma) \, \\ \, \sin(2 \phi_\gamma) \, \end{pmatrix}
    \label{eq:model_exy_gammaxy}
.\end{equation}

The parameters of the lens model are retrieved by minimising the chi-square:
\begin{equation}
\chi^2 = \chi_{\text{pos}}^2 + \chi_{\text{flux}}^2 = \sum_{i=1}^4 \begin{pmatrix} x_i - \hat{x}_i \\ y_i - \hat{y}_i \end{pmatrix}^T \mat{C}_i^{-1} \begin{pmatrix} x_i - \hat{x}_i \\ y_i - \hat{y}_i \end{pmatrix} + \frac{\left( f_i - \hat{\mu}_i \, f_s \right)^2}{\sigma_{f_i}^2}
\label{eq:model_chi2}
,\end{equation}
where $\mat{C}_i^{-1}$ is the inverse covariance matrix on $(x_i, y_i)$; $f_s$ is the original, non-lensed source flux; and $\sigma_{f_i}^2$ are the uncertainties on $f_i$.  If we assume that all lensed images have valid positions and fluxes, then Eq. \ref{eq:model_chi2} has twelve constraints and ten free parameters, yielding two degrees of freedom. Using particle swarm optimisation, we found that the minimum of Eq. \ref{eq:model_chi2}  \citep{Kennedy1995} first converges to an initial guess of the parameters, which were then refined using a burned-in Monte Carlo Markov chain (MCMC) sampler \citep{emcee} that simultaneously provides the uncertainties on the final parameters. The SIE+shear lens model also requires $f_s > 0$ and $e_x^2 + e_y^2 < 1,$ while we further impose $-0.5 < \gamma_x < 0.5$ and $-0.5 < \gamma_y < 0.5$ such that $\gamma < 2^{-\frac{1}{2}}$.

The relatively large positional chi-squares observed in Table \ref{tbl:lens_modelling_parameters} (median $\chi_{\text{pos}}^2$ = 72) can largely be explained by the very low astrometric uncertainty of {\it Gaia} (e.g. a median uncertainty of 0.27 mas in our sample) along with gravitational lensing effects that are not taken into account by the SIE+shear lens model: galaxy sub-structure, multiple deflectors, intervening galaxies along the line of sight, and micro-lensing. Micro-lensing, when combined with time delays and with the limitations of our model, is also the main cause of the large flux chi-squares (median $\chi_{\text{flux}}^2$ = 717). 

If we ignore the very low astrometric uncertainties of {\it Gaia}, most of the image positions are accurately reproduced by our model with a median root mean square (rms) error of $1.6$ mas according to Table \ref{tbl:lens_modelling_parameters}. Notable exceptions are:
\begin{itemize}
    \item[-] J1606-2333 (rms of 134 mas). Image C has loose constraints that permit a large rms with a reasonable $\chi_{\text{pos}}^2$ of $36.67$. The main deflector also has a companion object near image C that can potentially explain the observed discrepancy \citep{2019MNRAS.483.5649S}.
    \item[-]  GraL065904.1+162909 (rms of 84 mas). This system is known to be complicated to model with a SIE+shear lens model and has a satellite object close to image D \citep{2023MNRAS.518.1260S,2023A&A...672A...2E,2021ApJ...921...42S}.
    \item[-] B1422+231 (rms of 53 mas). Image D has large positional uncertainties of $\sigma_x=70$ mas and $\sigma_y=46$ mas, although its large $\chi_{\text{pos}}^2$ of $450.13$ comes from image B. \cite{2012A&A...538A..99S} had to introduce a second lensing galaxy to properly model this system.
    \item[-] RXJ0911+0551 (rms of 37 mas). Images A and C are inaccurately reproduced owing to the presence of a small satellite galaxy \citep{2000ApJ...544L..35K}.
    \item[-] GraL081828.3-261325 (rms of 32 mas). Images C and D are inaccurately reproduced as the deflector presumably consists of a group or cluster of galaxies \citep{2023MNRAS.518.1260S,2021ApJ...921...42S}.
    \item[-] 2MASSJ13102005-1714579 (rms of 22 mas). Images B and C are inaccurately reproduced as this system has two noticeable lensing galaxies (see Appendix \ref{charts_sec}).
\end{itemize}
All other systems have an rms of less than 7 mas. We should note here that the presence of multiple deflectors does not imply that the image positions are inaccurately reproduced since, for example, the cluster-lensed system SDSS1004+4112 has an rms of 0.2 mas. We finally compared the predicted position of the deflector to the position of the photo-centre provided in our catalogue for 11 systems and found a rather good agreement, of the order of $500$ mas, with the exception of SDSS1004+4112, where a shift of $1.4\arcsec$ is observed due to its wide separation and due to the fact that the deflector is composed of a cluster of galaxies. These differences, along with the discrepant values of the SIE+shear parameters we obtained compared to literature values, mainly come from the fact that we voluntarily do not artificially inflate the uncertainties on the observed fluxes, as is commonly done in lens modelling to compensate for the effect of micro-lensing, time delays, and galaxy sub-structures.

Accordingly, more sophisticated models exist in the literature that we do not aim to compete with. When available, these models should be preferred to the simple SIE+shear model presented here. A non-exhaustive list of such models can be found in \cite{2023MNRAS.518.1260S} for the quads with indices 1, 3, 4, 8, 13, 15, 17, and 18 in Table \ref{tbl:lens_modelling_parameters}; \cite{2023A&A...672A...2E} for 3, 10, 13, and 15; \cite{2021ApJ...921...42S} for 3, 5, 9, and 15; \cite{2020MNRAS.498.1440R} for 16; \cite{2019MNRAS.483.5649S} for 1, 14, and 18; \cite{2018MNRAS.476..927L} for 11; \cite{2017MNRAS.465.4895W} for 2; \cite{2012A&A...538A..99S} for 5, 7, and 12; \cite{2010A&A...522A..95C} for 2 and 10;  and \cite{2004AJ....128.2631W} for 6.

%***************************************************************************************************
\section{Conclusions}\label{ccl}
%***************************************************************************************************

We provide a comprehensive catalogue containing the most relevant information for each known lensed quasar and its components, including the best absolute astrometry, redshifts, and time delays available. The catalogue comprises 364 spectroscopically confirmed lensed systems, including 80 quadruply imaged quasars, 7 triply imaged quasars, and 277 doublets, along with precise astrometric data for 218 deflecting galaxies. The catalogue incorporates positional and photometric information from {\it Gaia} DR3 and the GravLens FPR, the best absolute positions currently available.

Redshifts for 347 quasars and 188 lensing galaxies have been compiled from the literature and major spectroscopic catalogues. We also include a curated set of 195 published individual time delays for 73  lensed systems. This information is essential for any further adoption of these data for time-delay cosmography, particularly in light of the forthcoming extended time-series data from {\it Gaia} DR4 (2026). 

A new important product of our catalogue is a list of published time delays that have been matched to the catalogue and correctly associated with the various components. 

We also provide simple but homogeneous lens modelling for 18 quads with complete astrometric and photometric coverage in {\it Gaia} DR3, using an SIE+external shear framework. While residuals highlight the limits of such simplified mass models, they nonetheless offer a robust baseline for lens configuration studies and training data for automated lens-finding algorithms.

The new catalogue serves a dual role: as a benchmark for validating machine-learning classifiers in lens searches, and as ready-to-use input for strong lensing models and cosmological applications.

%***************************************************************************************************
\begin{acknowledgements}
%***************************************************************************************************
We acknowledge support from `Observatoire Aquitain des Sciences de l'Univers (OASU)', from `Service National d'Observation (SNO) Gaia', and from `Actions sur projet INSU-PNGRAM'. The authors wish to acknowledge support from the ESA PRODEX Programme `Gaia-DPAC QSOs' and from the Belgian Federal Science Policy Office. The authors acknowledge support from the FAPESP (Fundação de Amparo à Pesquisa do Estado de São Paulo) and  S-Plus collaboration. This research has made use of the VizieR catalogue access tool, CDS, Strasbourg, France. The original description of the VizieR service was published in A\&AS 143, 23. This research has made use of ``Aladin sky atlas'' developed at CDS, Strasbourg Observatory, France. This work has made use of results from the ESA space mission {\it Gaia}, the data from which were processed by the {\it Gaia} Data Processing and Analysis Consortium (DPAC). Funding for the DPAC has been provided by national institutions, in particular the institutions participating in the {\it Gaia} Multilateral Agreement. The {\it Gaia} mission website is: http://www.cosmos.esa.int/gaia. Some of the authors are members of the {\it Gaia} Data Processing and Analysis Consortium (DPAC). This research is based on observations made with the NASA/ESA Hubble Space Telescope obtained from the Space Telescope Science Institute, which is operated by the Association of Universities for Research in Astronomy, Inc., under NASA contract NAS 5–26555. The Pan-STARRS1 Surveys (PS1) and the PS1 public science archive have been made possible through contributions by the Institute for Astronomy, the University of Hawaii, the Pan-STARRS Project Office, the Max-Planck Society and its participating institutes, the Max Planck Institute for Astronomy, Heidelberg and the Max Planck Institute for Extraterrestrial Physics, Garching, The Johns Hopkins University, Durham University, the University of Edinburgh, the Queen's University Belfast, the Harvard-Smithsonian Center for Astrophysics, the Las Cumbres Observatory Global Telescope Network Incorporated, the National Central University of Taiwan, the Space Telescope Science Institute, the National Aeronautics and Space Administration under Grant No. NNX08AR22G issued through the Planetary Science Division of the NASA Science Mission Directorate, the National Science Foundation Grant No. AST–1238877, the University of Maryland, Eotvos Lorand University (ELTE), the Los Alamos National Laboratory, and the Gordon and Betty Moore Foundation. This work is based [in part] on observations made with the NASA/ESA/CSA James Webb Space Telescope. The data were obtained from the Mikulski Archive for Space Telescopes at the Space Telescope Science Institute, which is operated by the Association of Universities for Research in Astronomy, Inc., under NASA contract NAS 5-03127 for JWST. DESI construction and operations is managed by the Lawrence Berkeley National Laboratory. This research is supported by the U.S. Department of Energy, Office of Science, Office of High-Energy Physics, under Contract No. DE–AC02–05CH11231, and by the National Energy Research Scientific Computing Center, a DOE Office of Science User Facility under the same contract. Additional support for DESI is provided by the U.S. National Science Foundation, Division of Astronomical Sciences under Contract No. AST-0950945 to the NSF’s National Optical-Infrared Astronomy Research Laboratory; the Science and Technology Facilities Council of the United Kingdom; the Gordon and Betty Moore Foundation; the Heising-Simons Foundation; the French Alternative Energies and Atomic Energy Commission (CEA); the National Council of Science and Technology of Mexico (CONACYT); the Ministry of Science and Innovation of Spain, and by the DESI Member Institutions. The DESI collaboration is honored to be permitted to conduct astronomical research on Iolkam Du’ag (Kitt Peak), a mountain with particular significance to the Tohono O’odham Nation. This research is based [in part] on data collected at the Subaru Telescope, which is operated by the National Astronomical Observatory of Japan. We are honored and grateful for the opportunity of observing the Universe from Maunakea, which has the cultural, historical, and natural significance in Hawaii.
\end{acknowledgements}

%%%%%%%%%%%%%%%%%%%%%%%%%%%%%%%%%%%%%%%%%%%%%%%%
\bibliographystyle{aa}
\bibliography{bibliography_dedup}

%%%%%%%%%%%%%%%%%%%%%%%%%%%%%%%%%%%%%%%%%%%%%%%%%%%%%%%%
% APPENDIX
%%%%%%%%%%%%%%%%%%%%%%%%%%%%%%%%%%%%%%%%%%%%%%%%%%%%%%%%
\newpage
\begin{appendix}
\onecolumn

%%%%%%%%%%%%%%%%%%%%%%%%%%%%%%%%%%%%%%%%%%%%%%%%
\section{Portion of the catalogue of multiply imaged quasars}\label{base}
%%%%%%%%%%%%%%%%%%%%%%%%%%%%%%%%%%%%%%%%%%%%%%%%
 \fontsize{8.7}{10}\selectfont 
 \begin{longtable}{rlrrrrrrl}
 \caption{\label{catalogue} Portion of the catalogue of known lenses presenting quadruply imaged quasars fully measured by \textit{Gaia} DR3 and a selection of properties from the catalogue. } \\
 \hline
 Index& Name & Comp. & \textit{Gaia} DR3 source\_id & z$_s$ & z$_l$ & G [mag] & $G_{\mathrm{BP}}$ [mag] & $G_{\mathrm{RP}}$ [mag]\\
 \hline
1& 2MASXJ01471020+4630433  &  A  &  350937280928094336  &  2.377  &  0.678  &  15.944  &  15.69  &  14.83\\
1& 2MASXJ01471020+4630433  &  B  &  350937280925970432  &  2.377  &  0.678  &  16.717  &  16.32  &  15.36\\
1& 2MASXJ01471020+4630433  &  C  &  350937280925970304  &  2.377  &  0.678  &  16.173  &  16.29  &  15.54\\
1& 2MASXJ01471020+4630433  &  D  &  350937280925971456  &  2.377  &  0.678  &  18.300  &  18.38  &  17.60\\
\hline
\multicolumn{9}{l}{\tablefoot{The online version of the catalogue also includes the astrometry and photometry of \textit{Gaia} DR3, the FPR data, the AllWISE photometry, redshifts from Milliquas, \textit{Gaia} QSOC DR3 or SHSRC when available and the bibliographic references. }}
\end{longtable}
The references we searched for discoveries of lenses are: 
  \cite{
2015MNRAS.454.1260A, 2018MNRAS.474.3391A, 2018MNRAS.479.4345A, 2018MNRAS.475.2086A, 2009A&A...507...35A,  
2018MNRAS.480.5017A, 2003MNRAS.338..957A, 2001MNRAS.326.1007A, 1997A&A...317L..13B, 1992AJ....103..405B,  
2017ApJ...844...90B, 2003MNRAS.338.1084B, 2008AJ....135..374B, 1993MNRAS.263L..32B, 2006ApJ...652..955C,  
2022A&A...659A.140C, 2010A&A...522A..95C, 1997A&A...318L..67C, 2022ApJ...925..162C, 1996A&A...305L...9C,  
2002A&A...387..406C, 2013ApJ...773..146D, 2023ApJS..269...61D, 2019A&A...622A.165D, 2022MNRAS.509..738D,  
2023A&A...679L...4D, 2024A&A...682A..47D, 2006A&A...451..759E, 2019ApJ...870L..11F, 1998AJ....115..377F,  
1999AJ....117..658F, 1998ApJ...503L.127F, 2023MNRAS.522.5142F, 2025A&A...698A.227F, 2023ApJ...943...25G,  
2016A&A...596A..77G, 2023ApJ...956..117G, 2025ApJ...989..112G, 2000A&A...357L..29H, 2021MNRAS.503.3848H,  
2025arXiv250903858H, 2025arXiv250108541H, 1992AJ....104..968H, 1985AJ.....90..691H, 2003Natur.426..810I,  
2003AJ....126..666I, 2005AJ....130.1967I, 2006ApJ...653L..97I, 2006AJ....131.1934I, 2007AJ....133..206I,  
2012AJ....143..119I, 2014AJ....147..153I, 1998ApJ...505..529I, 1995MNRAS.274L..25J, 2008MNRAS.387..741J,  
2012MNRAS.419.2014J, 2021MNRAS.502.1487J, 2024BSRSL..93..752J, 2003AJ....126.2281J, 2010AJ....139.1614K,  
1997ApJ...479..678K, 1999MNRAS.303..727K, 2018MNRAS.476..663K, 2019arXiv191208977K, 2018A&A...616L..11K,  
2002AJ....123.2925L, 1989AJ.....97.1283L, 1997AJ....114...48L, 2019MNRAS.483.4242L, 2020MNRAS.494.3491L,  
2018MNRAS.479.5060L, 2023MNRAS.520.3305L, 2023ApJ...955L..16L, 2017ApJ...838L..15L, 2000AJ....119..451L,  
2018MNRAS.476..927L, 1988Natur.334..325M, 1999AJ....118..654M, 2001AJ....121..619M, 2023ApJ...946...63M,  
2004MNRAS.350..167M, 2005MNRAS.356.1009M, 1992Gemin..36....1M, 2016MNRAS.456.1595M, 2017MNRAS.465.2411M,  
1999AJ....118.1444M, 2001AJ....121..611M, 2003AJ....126..696M, 2003AJ....126.2145M, 2007AJ....133..214M,  
1995ApJ...447L...5M, 1999AJ....117.2565M, 2023ApJ...954L..38N, 1984ApJ...283..512N, 2004PASJ...56..399O,  
2005ApJ...622..106O, 2017MNRAS.465.4325O, 1992MNRAS.259P...1P, 2000MNRAS.319L...7P, 2004AJ....127.1318P,  
2006AJ....131...41P, 1988MNRAS.231..229P, 1995ApJ...453L...5R, 1999AJ....117.2010R, 2002A&A...382L..26R,  
2016MNRAS.458....2R, 1998AJ....115.1371S, 2017AJ....153..219S, 2018RNAAS...2...21S, 2024arXiv240414256S,  
2023MNRAS.518.1260S, 2024A&A...690A..57S, 2019MNRAS.483.5649S, 2021MNRAS.503.1557S, 2018MNRAS.480.2853S,  
2018MNRAS.481L.136S, 2019MNRAS.489.4741S, 2003A&A...406L..43S, 2019MNRAS.483.3888S, 2019MNRAS.485.5086S,  
2021ApJ...921...42S, 1987Natur.329..695S, 1993LIACo..31..153S, 2023A&A...672L...9T, 1979Natur.279..381W,  
2017MNRAS.468.3757W, 1980Natur.285..641W, 2018MNRAS.477L..70W, 2000AJ....120.2868W, 2001AJ....121.1223W,  
2002AJ....123...10W, 2002ApJ...564..143W, 1993A&A...278L..15W, 1996A&A...315L.405W, 1999A&A...348L..41W,  
2002A&A...395...17W, 2004A&A...419L..31W, 2003A&A...405..445W, 2005PhDT.......113Y, 2023AJ....165..191Y}. 

The reference we searched for spectroscopic redshifts are: \cite{1985AJ.....90..691H, 1986AJ.....91..991S, 1995A&A...297L..59W, 1996A&A...307L..53C, 1996ApJ...460L.103F, 1996Natur.379..139W, 1998ApJ...503..118S, 1998MNRAS.301..310S, 1999AJ....117..658F, 1999AJ....117.2034T, 2000A&A...364L..62L, 2000AJ....119..451L, 2000AJ....119.1078T, 2000AJ....120.2868W, 2001AJ....121..611M, 2001ApJ...557..594R, 2002A&A...382L..26R, 2002A&A...387..406C, 2002A&ARv..10..263C, 2002AJ....123.2925L, 2003AJ....126.2145M, 2003MNRAS.338.1084B, 2003Natur.426..810I, 2005MNRAS.356.1009M, 2006A&A...451..747E, 2006A&A...451..759E, 2006AJ....131.1934I, 2006ApJ...641...70O, 2006ApJ...652..955C, 2006ApJ...653L..97I, 2007A&A...465...51E, 2007AJ....133..214M, 2008AJ....135..520O, 2009A&A...507...35A, 2010A&A...516L..12C, 2010A&A...518A..10V, 2010AJ....139.1614K, 2010AJ....140..370M, 2011ApJ...730..108R, 2011ApJ...738...30R, 2012A&A...538A..99S, 2012AJ....143..119I, 2012MNRAS.419.2014J, 2013ApJ...765..139R, 2013ApJ...773..146D, 2014AJ....147..153I, 2015MNRAS.454.1260A, 2016MNRAS.456.1595M, 2017AJ....153..219S, 2017ApJ...838L..15L, 2017MNRAS.465.2411M, 2017MNRAS.468.3757W, 2018A&A...616A.118G, 2018MNRAS.474.3391A, 2018MNRAS.475.2086A, 2018MNRAS.476..663K, 2018MNRAS.476..927L, 2018MNRAS.477L..70W, 2018MNRAS.479.5060L, 2018MNRAS.480.5017A, 2018MNRAS.481L.136S, 2018RNAAS...2...21S, 2023ApJ...943...25G, 2019ApJ...870L..11F, 2019MNRAS.483.4242L, 2019MNRAS.485.5086S, 2019MNRAS.486.4987R, 2019MNRAS.489.4741S, 2019arXiv191208977K, 2020A&A...642A.193M, 2020MNRAS.494.3491L, 2021AJ....162..175O, 2021ApJ...921...42S, 2021MNRAS.502.1487J, 2021MNRAS.503.3848H, 2022A&A...659A.140C, 2022MNRAS.509..738D, 2023A&A...672A..20M, 2023A&A...672L...9T, 2023Coryn, 2023A&A...679L...4D, 2023AJ....165..191Y, 2023ApJ...946...63M, 2023ApJ...954L..38N, 2023ApJ...955L..16L, 2023ApJ...956..117G, 2023ApJS..269...61D, 2023MNRAS.520.3305L, 2023MNRAS.522.5142F, 2023OJAp....6E..49F, 2024A&A...682A..47D, 2024A&A...689A.129F, 2024A&A...690A..57S, 2024MNRAS.530..221J, 2024MNRAS.535.1652K, 2024arXiv240803379C, 2024arXiv240916363G, 2024ApJ...976..110A, 2024arXiv241104177D, 2025arXiv250108541H}.

%%%%%%%%%%%%%%%%%%%%%%%%%%%%%%%%%%%%%%%%%%%%%%%%
%%%%%%%%%% TABLE TIME_DELAYS        %%%%%%\label{timedelay}
%%%%%%%%%%%%%%%%%%%%%%%%%%%%%%%%%%%%%%%%%%%%%%%%
%\newpage
\section{Portion of the catalogue of published time delays}\label{delays}
%%%%%%%%%%%%%%%%%%%%%%%%%%%%%%%%%%%%%%%%%%%%%%%%

\fontsize{8}{10}\selectfont % \fontsize{8.7}{10}\selectfont
\begin{longtable}{lrrrrrrl}
\caption{\label{timedelay_table}Time delays of known gravitational lens systems as collected from the literature. } \\

\hline
Name & $\Delta$$t_{AB}$ & $\Delta$$t_{AC}$ & $\Delta$$t_{AD}$ & $\Delta$$t_{BC}$ & $\Delta$$t_{BD}$ & $\Delta$$t_{CD}$ & Author \\
\multicolumn{8}{c}{[days]}\\
\hline
  DESJ0029-3814 & $-6.5 \pm 3.0$ & $-6.6 \pm 2.7$ & $43.1 \pm 2.4$ & $-2.0 \pm 3.5$ & $46.7 \pm 2.9$ & $49.9 \pm 2.7$ & 2025Dux\\
  PS J0030-1525* & $9.3 \pm 1.6$ & $-19.3 \pm 3.3$ &  & $-28.5 \pm 3.5$ &  &  & 2025Dux\\
  HE0047-1756 & $-10.8 \pm 1.0$ &  &  &  &  &  & 2020Millon\_b\\
  DES J0053-2012 & $-26.7 \pm 2.6$ & $-20.2 \pm 2.6$ & $-90.2 \pm 6.7$ & $6.3 \pm 2.0$ & $-63.5 \pm 5.9$ & $-70.2 \pm 6.1$ & 2025Dux\\
  Q0142-100 & $-97.7 \pm 16.1$ &  &  &  &  &  & 2020Millon\_a\\
\multicolumn{8}{l}{\tablefoot{The A, B, C, and D components refer to the lens catalogue, which usually uses the discovery paper identification. Sources marked with a (*) sign correspond to cases where two components were not separated during time delay measurements; see original publication for details.}}\\

\end{longtable}
The references we searched for time delays are: 
\cite{2025Dux,2020Millon_a,2020Millon_b,2018MNRAS.476.5393B,
2018A&A...609A..71C, 2023A&A...672L...9T, 2023A&A...674A..24C,
2024AstBu..79...15B, 2016A&A...587A..43S, 2013ApJ...774...69H, 
2025A&A...694A..31S, 2008A&A...492..401S, 2017ApJ...834...31A, 
2022ApJ...937...34M, 2013ApJ...764..186F, 2018MNRAS.481.1000B, 
2008ApJ...676...80M, 2018A&A...616A.183B, 2013A&A...553A.121E, 
2016A&A...596A..77G, 2017MNRAS.465.3607A, 2001MNRAS.326.1403P, 
2023arXiv231209311Q, 2019ApJ...873..117S, 2023ApJ...950...37M, 
2011A&A...536A..44E, 2021MNRAS.505.2610B, 2002ApJ...581..823F, 
2017MNRAS.468.3757W, 2024ApJ...964..173R, 2023A&A...674A.101M, 
2019A&A...629A..97B, 2021MNRAS.503.3848H, 2015ApJ...813...67D,
2026A&A...706L..12N, 2026MNRAS.546ag023B}.

%%%%%%%%%%%%%%%%%%%%%%%%%%%%%%%%%%%%%%%%%%%%%%%%
%%%%%%%%%% TABLE MODELLING        %%%%%%
%%%%%%%%%%%%%%%%%%%%%%%%%%%%%%%%%%%%%%%%%%%%%%%%
\onecolumn
\begin{landscape}
\section{Parameters of the SIE+shear lens model for quads with four detections in \textit{Gaia} DR3}
\label{sec:lens_modelling_parameters}
{ \scriptsize
\setlength{\tabcolsep}{4pt} % Default value: 6pt
\renewcommand{\arraystretch}{2} % Default value: 1
%\begin{longtable}{lcccccccccccccc}
\begin{longtable}{rlrrrrrrrlcccccccccc}
\caption{Parameters of the SIE+shear lens model for the 18 quadruply imaged quasars having four lensed images available in {\it \textit{Gaia}} DR3. }\label{tbl:lens_modelling_parameters} \\
\hline
Index & Name & $\alpha_1$ & $\delta_1$ & $\chi_{\text{pos}}^2$ & $\chi_{\text{flux}}^2$ & $\text{rms}_{\text{pos}}$ & $\sigma_{\text{pos}}$ & $\text{S/N}_{\text{flux}}$ & $\theta_E$ & $e_x$ & $e_y$ & $\gamma_x$ & $\gamma_y$ & $x_g$ & $y_g$ & $d_g$ & $x_s$ & $y_s$ & $f_s$ \\
& & [degree] & [degree] & & & [$\mu$as] & [$\mu$as] & & [arcsec] & & & & & [arcsec] & [arcsec] & [mas] & [arcsec] & [arcsec] & [e$^-$ s$^{-1}$] \\ \hline
1& 2MASXJ01471020+4630433 & 26.7924402 & 46.5121697 & 188.50 & 17392.84 & 1654 & 159 & 323 & $1.9061$ & $0.2920$ & $-0.0981$ & $0.2466$ & $-0.07451$ & $-0.2165$ & $-1.7890$ & 543 & $-0.2728$ & $-2.0261$ & $240.2$\\
2& HE0435-1223 & 69.5623046 & -12.2873141 & 4.95 & 375.70 & 150 & 116 & 139 & $1.2040$ & $0.081$ & $0.132$ & $0.0809$ & $0.102$ & $-1.1901$ & $-0.5395$ & 177 & $-1.0422$ & $-0.447$ & $33.7$\\
3& GraLJ065904.1+162909 & 104.7673183 & 16.4861288 & 250.57 & 2091.74 & 84721 & 531 & 128 & $3.21$ & $-0.589$ & $-0.535$ & $-0.151$ & $-0.116$ & $-1.282$ & $-0.773$ & & $-2.827$ & $-1.294$ & $76.6$\\
4& GraLJ081828.3-261325 & 124.6185812 & -26.2233743 & 3060.71 & 1296.93 & 32312 & 558 & 293 & $2.9444$ & $-0.1801$ & $0.201$ & $-0.1395$ & $0.227$ & $-2.8362$ & $-0.835$ & & $-2.991$ & $-0.350$ & $50.1$\\
5& RXJ0911+0551 & 137.8651357 & 5.8484102 & 136.57 & 242.83 & 37858 & 1328 & 73 & $1.16$ & $-0.49$ & $-0.07$ & $0.22$ & $-0.110$ & $-0.74$ & $0.1424$ & 50 & $-1.24$ & $0.09$ & $148 \pm 2$\\
6& SDSS1004+4112 & 151.1450314 & 41.2108979 & 4.63 & 590.19 & 209 & 256 & 101 & $7.7683$ & $-0.2266$ & $0.2580$ & $-0.12519$ & $0.2229$ & $-6.1117$ & $3.408$ & 1447 & $-7.4124$ & $3.854$ & $16.4$\\
7& PG1115+080 & 169.5706675 & 7.7662505 & 18.09 & 312.31 & 423 & 218 & 128 & $1.1477$ & $0.090$ & $-0.007$ & $-0.023$ & $-0.101$ & $-1.1167$ & $0.231$ & 232 & $-1.085$ & $0.248$ & $202 \pm 4$\\
8& GraLJ113100013-441959935 & 172.7503126 & -44.3332479 & 72.00 & 359.01 & 6507 & 923 & 83 & $0.8547$ & $-0.082$ & $0.054$ & $0.006$ & $0.026$ & $-0.778$ & $-0.410$ & & $-0.790$ & $-0.353$ & $21.1$\\
9& RXJ1131-1231 & 172.9649269 & -12.5330233 & 31.01 & 1617.67 & 826 & 190 & 66 & $1.850$ & $-0.248$ & $0.120$ & $-0.197$ & $0.1017$ & $-1.684$ & $0.431$ & 384 & $-1.917$ & $0.4760$ & $22.9$\\
10& 2MASSJ11344050-2103230 & 173.6691019 & -21.0564343 & 51.79 & 17660.11 & 639 & 63 & 317 & $1.328$ & $-0.0778$ & $-0.314$ & $-0.0143$ & $0.2372$ & $-0.7965$ & $0.8117$ & & $-0.8599$ & $0.7640$ & $1333 \pm 4$\\
11& 2MASSJ13102005-1714579\footnote{Image with {\tt source\_id}=3511426761399771776 has no valid {\tt phot\_g\_mean\_flux} in \textit{Gaia} DR3 such that we assumed {\tt phot\_g\_mean\_flux\_error}=$\infty$. Also, as 2MASSJ13102005-1714579 has two lensing galaxies, $d_g$ was not computed for this system.} & 197.5840297 & -17.2500719 & 73.76 & 717.43 & 22768 & 454 & 140 & $2.943$ & $-0.092$ & $-0.017$ & $-0.034$ & $0.024$ & $-1.663$ & $2.429$ & & $-1.764$ & $2.293$ & $13.3$\\
12& B1422+231 & 216.1588259 & 22.9335778 & 450.13 & 84.71 & 53301 & 685 & 379 & $0.898$ & $0.316$ & $-0.538$ & $0.124$ & $-0.048$ & $0.497$ & $-0.962$ & 148 & $0.025$ & $-0.8151$ & $1495 \pm 4$\\
13& J1606-2333 & 241.5012285 & -23.5559591 & 36.67 & 47.51 & 134155 & 600 & 115 & $0.7313$ & $0.138$ & $0.495$ & $0.119$ & $0.377$ & $-0.8132$ & $-0.3218$ & 82 & $-0.593$ & $-0.050$ & $45.4$\\
14& GraLJ165105.3-041725 & 252.7714043 & -4.2910582 & 1.38 & 4303.47 & 138 & 223 & 205 & $3.484$ & $0.0777$ & $0.371$ & $0.0920$ & $0.4167$ & $3.5681$ & $3.0329$ & & $2.169$ & $1.911$ & $83.9$\\
15& J1721+8842 & 260.4541941 & 88.7062147 & 87.16 & 7136.70 & 1808 & 148 & 188 & $2.188$ & $-0.494$ & $-0.3445$ & $-0.2629$ & $-0.1897$ & $-1.397$ & $-0.6307$ & 480 & $-2.177$ & $-0.8210$ & $41.9$\\
16& WFI2033-4723 & 308.4253501 & -47.3953751 & 26.11 & 3398.30 & 546 & 177 & 197 & $1.1437$ & $0.3364$ & $-0.1283$ & $0.2668$ & $-0.1388$ & $0.7928$ & $-0.8741$ & 177 & $0.5225$ & $-0.9781$ & $62.3$\\
17& GraLJ203802-400815 & 309.5116226 & -40.1374099 & 2.77 & 92.44 & 1508 & 442 & 134 & $1.366$ & $-0.017$ & $0.085$ & $0.029$ & $-0.057$ & $-0.858$ & $1.197$ & & $-0.845$ & $1.114$ & $38.2$\\
18& J2145+6345 & 326.2717031 & 63.7613612 & 262.95 & 7249.86 & 1140 & 159 & 349 & $1.1205$ & $0.3081$ & $0.5521$ & $0.1603$ & $0.3291$ & $-0.6954$ & $0.4092$ & & $-0.7971$ & $0.8245$ & $122.5$\\
\hline
\multicolumn{20}{l}{\tablefoot{Source and galaxy positions $(x_s, y_s)$ and $(x_g, y_g)$ are relative to the position of the brightest lensed component $(x_1, y_1) = 3600 \cdot ( \alpha_1 \cos \delta_1, \delta_1 )$. The combined chi-square, $\chi^2 = \chi_{\text{pos}}^2 + \chi_{\text{flux}}^2$ from Eq. \ref{eq:model_chi2}, has two degrees of freedom for all systems, except for 2MASSJ13102005-1714579 where one $G$-band flux is missing that leads to one degree of freedom. The root mean square error on the position and median positional uncertainty, both projected on the plane tangential to the celestial sphere, are respectively reported as $\text{rms}_\text{pos}$ and $\sigma_{\text{pos}}$, while the median S/N of the fluxes is reported as $\text{S/N}_{\text{flux}} \equiv \operatorname{median} \frac{f_i}{\sigma_{f_i}}$. Although the galaxy position is left as a free parameter, eleven systems have at least one galaxy position that is reported in our catalogue. For systems with a single lensing galaxy, $d_g$ is the distance between the predicted and observed galaxy position. Other parameters are described in Sect. \ref{model}. Parameters are truncated to the last significant decimal according to the MCMC 0.16 and 0.84 uncertainty percentiles or are explicitly written if the uncertainty is larger than one. As an example, a tabulated value of $\theta_E$ = 1.328 means that its uncertainty stands between 0.001 (inclusive) and 0.01 (exclusive).}}
\end{longtable}
}
\end{landscape}

%%%%%%%%%%%%%%%%%%%%%%%%%%%%%%%%%%%%%%%%%%%%%%%%
\newpage
\section{Finding charts of the quadruply and triply imaged quasars}\label{charts_sec}
%%%%%%%%%%%%%%%%%%%%%%%%%%%%%%%%%%%%%%%%%%%%%%%%

%%%%%%%%%%%%%%%%%%%%%%%%%%%%%%%%%%%%%%%%%%%%%%%%
%%%%%%%%%% FIGURE CHARTS ALL QUADS        %%%%%%\label{charts}
%%%%%%%%%%%%%%%%%%%%%%%%%%%%%%%%%%%%%%%%%%%%%%%%
\vspace{-3pt}
\begin{figure*}[h!]
\centering
\caption{\label{charts}
Finding charts for the 80 known quads and 7 triple systems plotted on top of JWST/F150W (J), HST (H -- with subscripted bands H, I, J, Y, wideV), Hyper Suprime-Cam/Subaru (S), Pan-STARRS (P), DESI Legacy Survey DR10 (L), DSS colored (D), Beijing-Arizona Sky Survey (B), CFHT Megacam (CM), or Keck 2 OSIRIS (O). Mean coordinates of the systems are indicated. The Identification of the A, B, C, D components and galaxies usually refers to the discovery publication. "S" stands for systems with only one published component. North is up, east is to the left.}
\includegraphics[width=0.16255\textwidth]{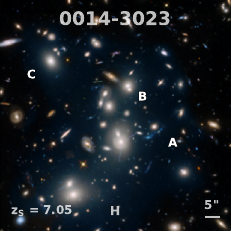}
\includegraphics[width=0.16255\textwidth]{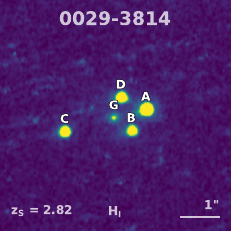}
\includegraphics[width=0.16255\textwidth]{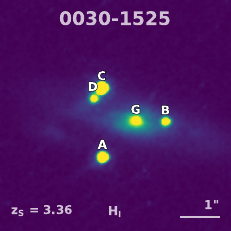}
\includegraphics[width=0.16255\textwidth]{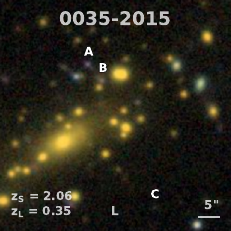}
\includegraphics[width=0.16255\textwidth]{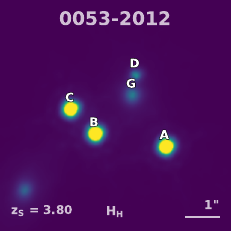}
\includegraphics[width=0.16255\textwidth]{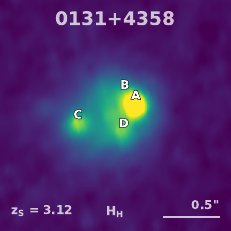}
\includegraphics[width=0.16255\textwidth]{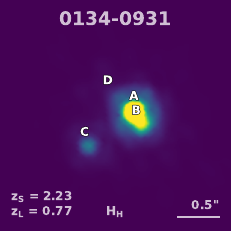}
\includegraphics[width=0.16255\textwidth]{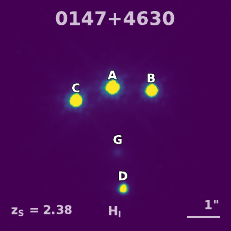}
\includegraphics[width=0.16255\textwidth]{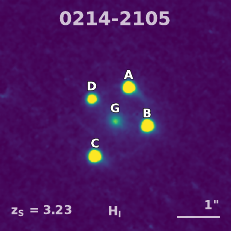}
\includegraphics[width=0.16255\textwidth]{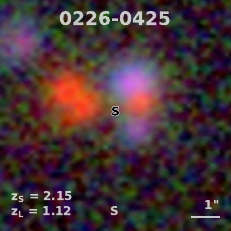}
\includegraphics[width=0.16255\textwidth]{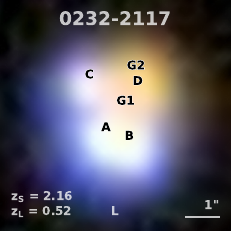}
\includegraphics[width=0.16255\textwidth]{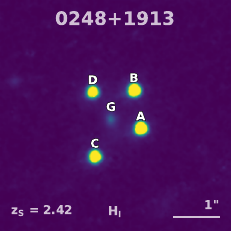}
\includegraphics[width=0.16255\textwidth]{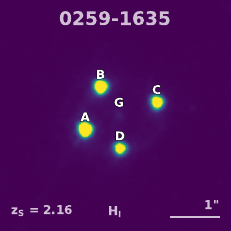}
\includegraphics[width=0.16255\textwidth]{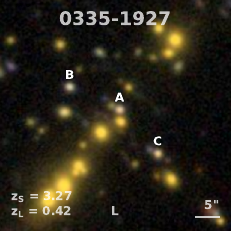}
\includegraphics[width=0.16255\textwidth]{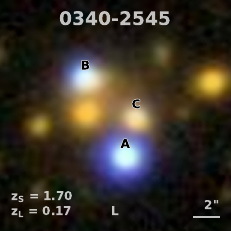}
\includegraphics[width=0.16255\textwidth]{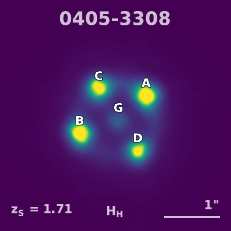}
\includegraphics[width=0.16255\textwidth]{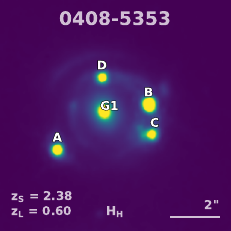}
\includegraphics[width=0.16255\textwidth]{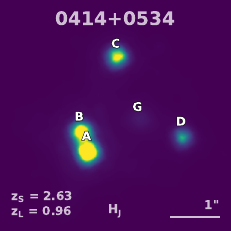}
\includegraphics[width=0.16255\textwidth]{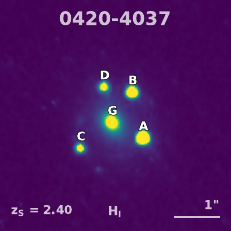}
\includegraphics[width=0.16255\textwidth]{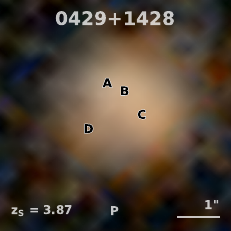}
\includegraphics[width=0.16255\textwidth]{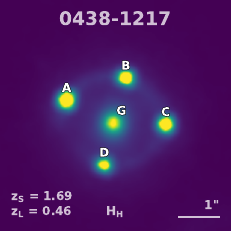}
\includegraphics[width=0.16255\textwidth]{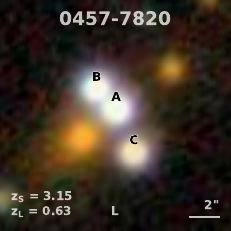}
\includegraphics[width=0.16255\textwidth]{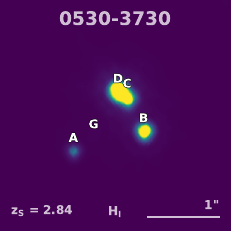}
\includegraphics[width=0.16255\textwidth]{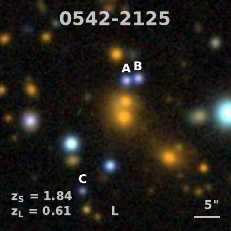}
\includegraphics[width=0.16255\textwidth]{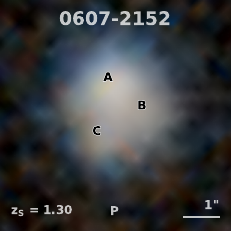}
\includegraphics[width=0.16255\textwidth]{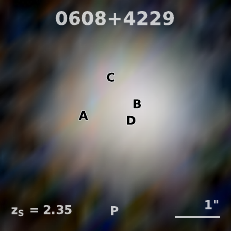}
\includegraphics[width=0.16255\textwidth]{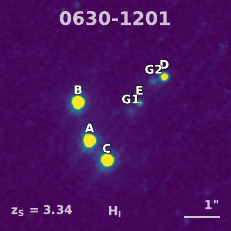}
\includegraphics[width=0.16255\textwidth]{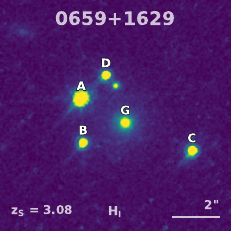}
\includegraphics[width=0.16255\textwidth]{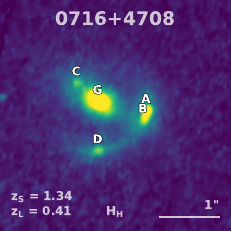}
\includegraphics[width=0.16255\textwidth]{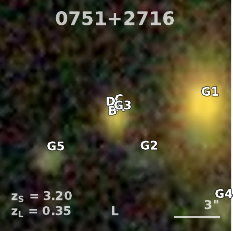}
\includegraphics[width=0.16255\textwidth]{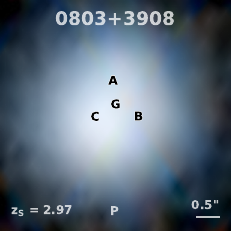}
\includegraphics[width=0.16255\textwidth]{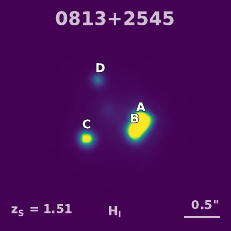}
\includegraphics[width=0.16255\textwidth]{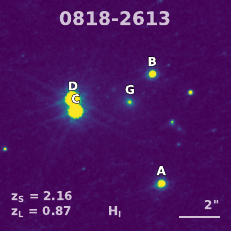}
\includegraphics[width=0.16255\textwidth]{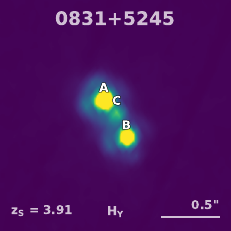}
\includegraphics[width=0.16255\textwidth]{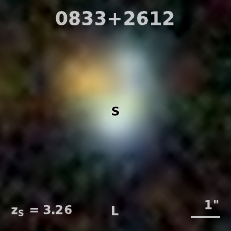}
\includegraphics[width=0.16255\textwidth]{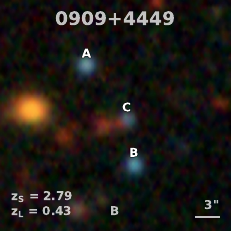}
\includegraphics[width=0.16255\textwidth]{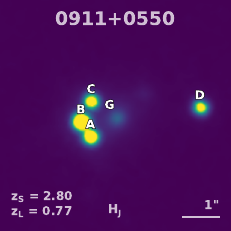}
\includegraphics[width=0.16255\textwidth]{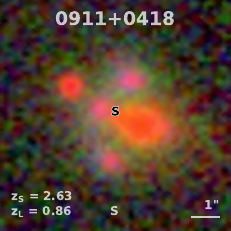}
\includegraphics[width=0.16255\textwidth]{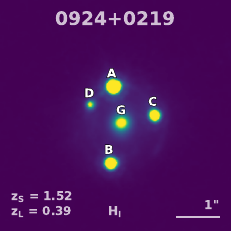}
\includegraphics[width=0.16255\textwidth]{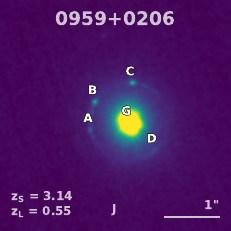}
\includegraphics[width=0.16255\textwidth]{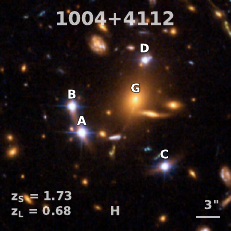}
\includegraphics[width=0.16255\textwidth]{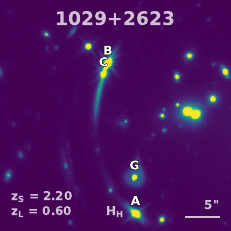}
\end{figure*}
\begin{figure*}[h!]
\centering
\includegraphics[width=0.16255\textwidth]{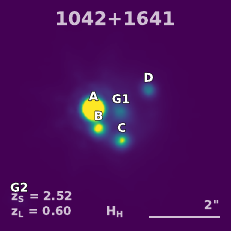}
\includegraphics[width=0.16255\textwidth]{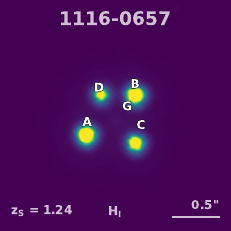}
\includegraphics[width=0.16255\textwidth]{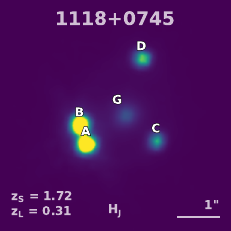}
\includegraphics[width=0.16255\textwidth]{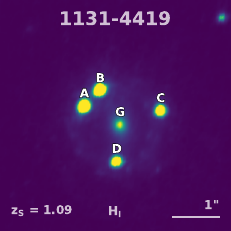}
\includegraphics[width=0.16255\textwidth]{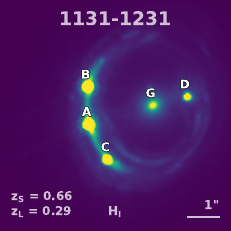}
\includegraphics[width=0.16255\textwidth]{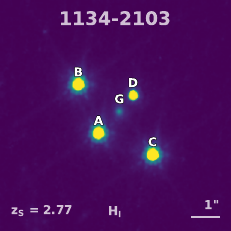}
\includegraphics[width=0.16255\textwidth]{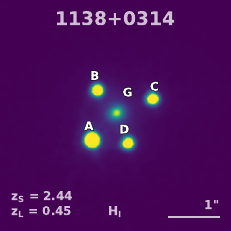}
\includegraphics[width=0.16255\textwidth]{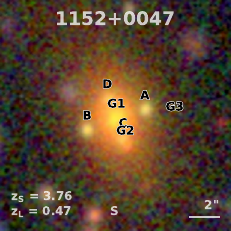}
\includegraphics[width=0.16255\textwidth]{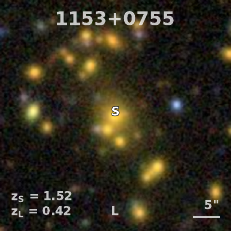}
\includegraphics[width=0.16255\textwidth]{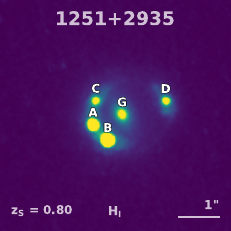}
\includegraphics[width=0.16255\textwidth]{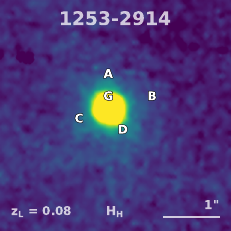}
\includegraphics[width=0.16255\textwidth]{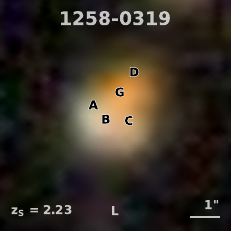}
\includegraphics[width=0.16255\textwidth]{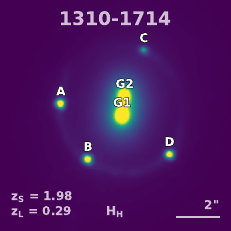}
\includegraphics[width=0.16255\textwidth]{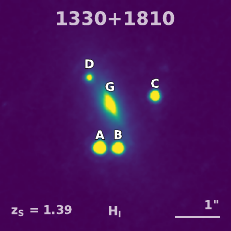}
\includegraphics[width=0.16255\textwidth]{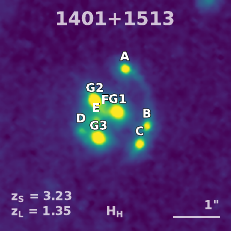}
\includegraphics[width=0.16255\textwidth]{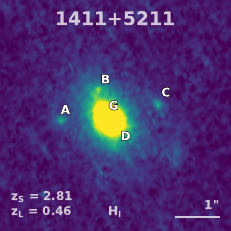}
\includegraphics[width=0.16255\textwidth]{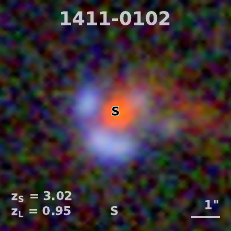}
\includegraphics[width=0.16255\textwidth]{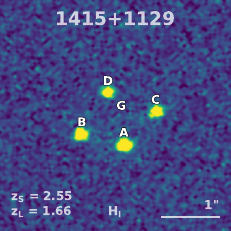}
\includegraphics[width=0.16255\textwidth]{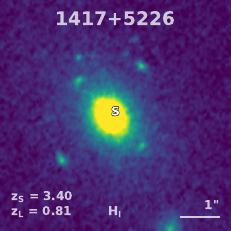}
\includegraphics[width=0.16255\textwidth]{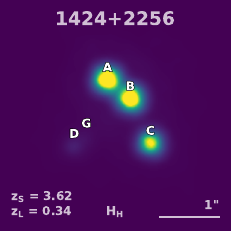}
\includegraphics[width=0.16255\textwidth]{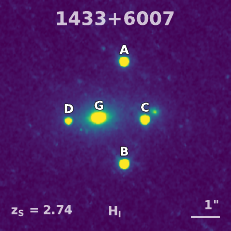}
\includegraphics[width=0.16255\textwidth]{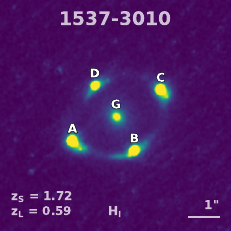}
\includegraphics[width=0.16255\textwidth]{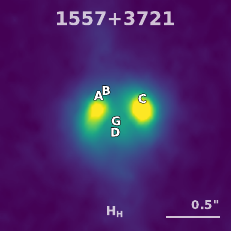}
\includegraphics[width=0.16255\textwidth]{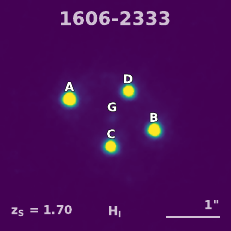}
\includegraphics[width=0.16255\textwidth]{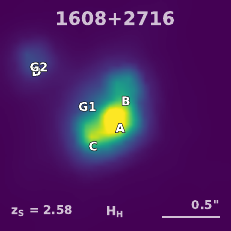}
\includegraphics[width=0.16255\textwidth]{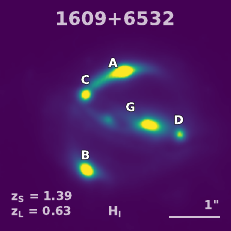}
\includegraphics[width=0.16255\textwidth]{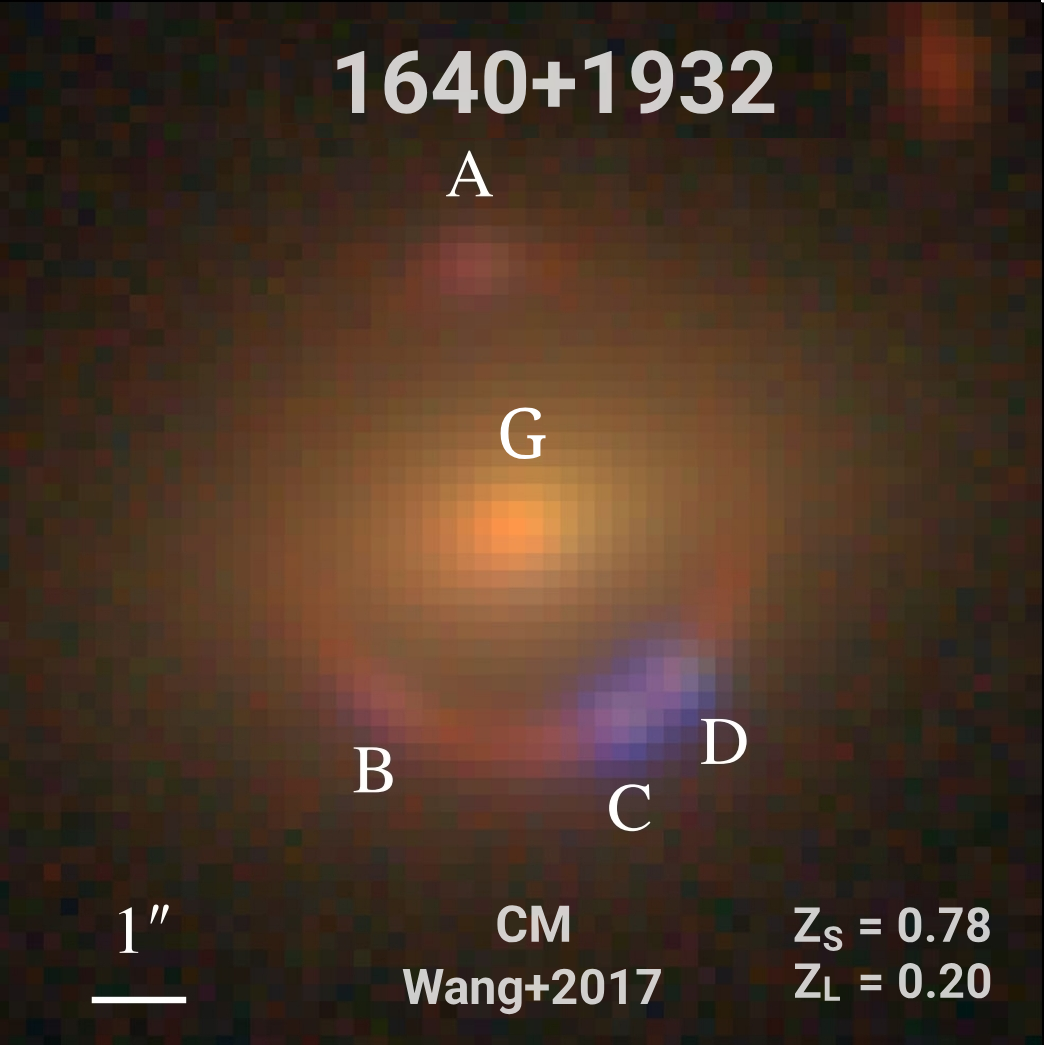}
\includegraphics[width=0.16255\textwidth]{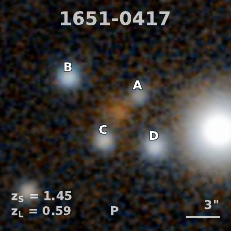}
\includegraphics[width=0.16255\textwidth]{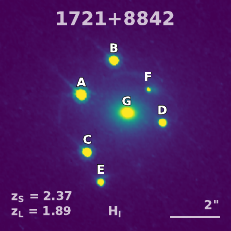}
\includegraphics[width=0.16255\textwidth]{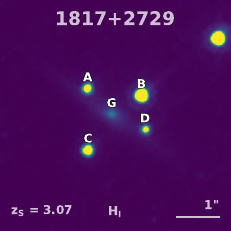}
\includegraphics[width=0.16255\textwidth]{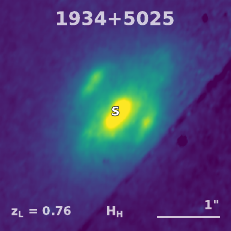}
\includegraphics[width=0.16255\textwidth]{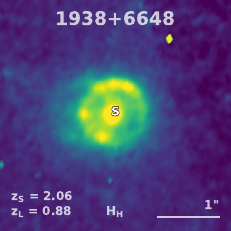}
\includegraphics[width=0.16255\textwidth]{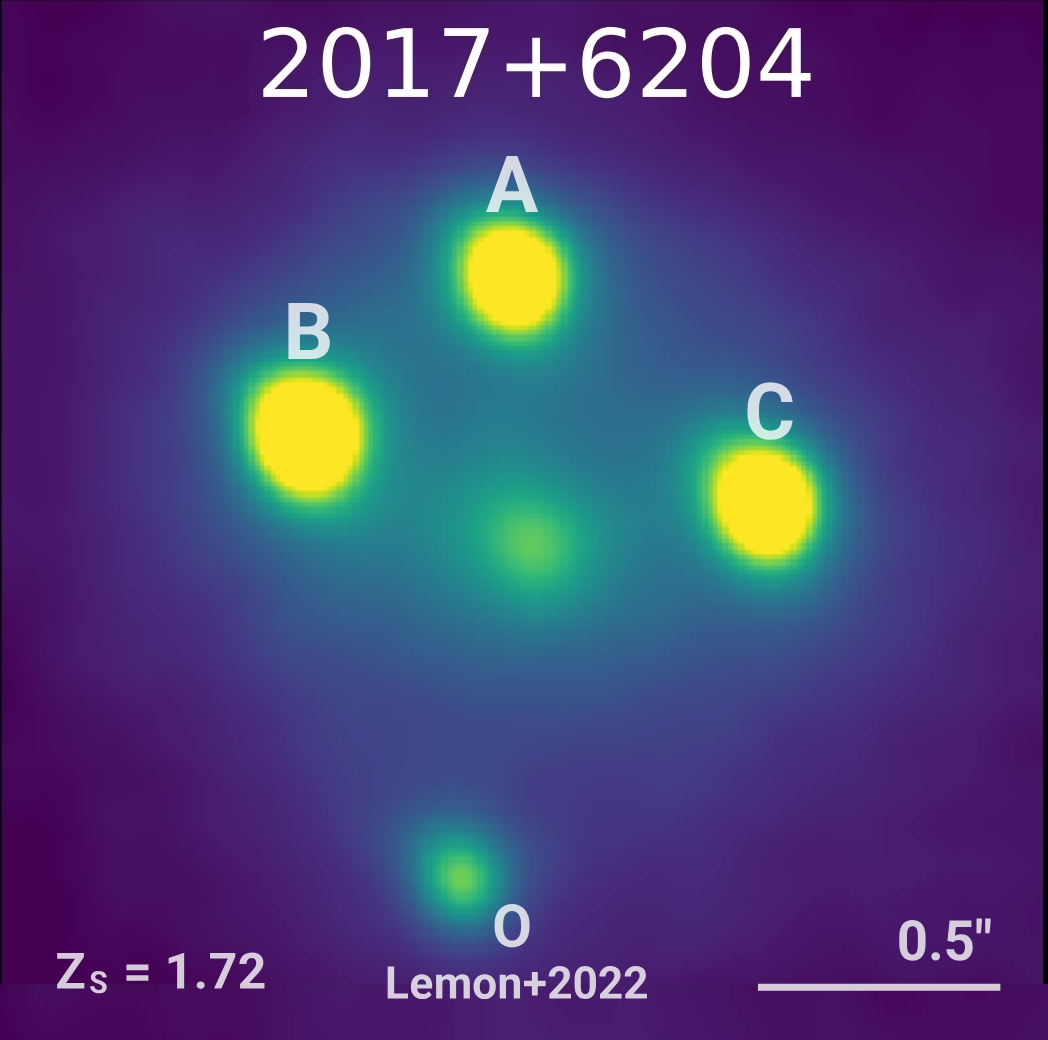}
\includegraphics[width=0.16255\textwidth]{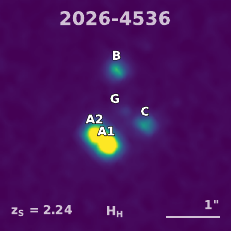}
\includegraphics[width=0.16255\textwidth]{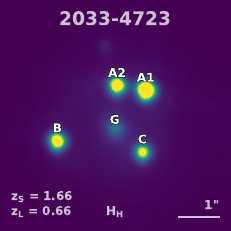}
\includegraphics[width=0.16255\textwidth]{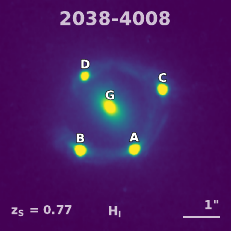}
\includegraphics[width=0.16255\textwidth]{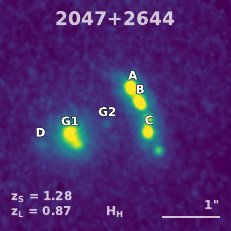}
\includegraphics[width=0.16255\textwidth]{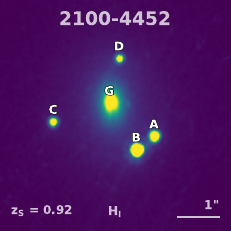}
\includegraphics[width=0.16255\textwidth]{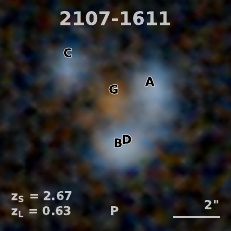}
\includegraphics[width=0.16255\textwidth]{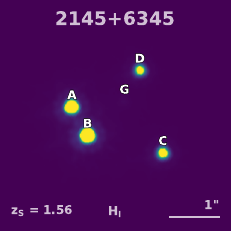}
\includegraphics[width=0.16255\textwidth]{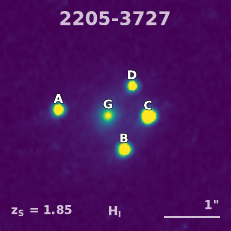}
\includegraphics[width=0.16255\textwidth]{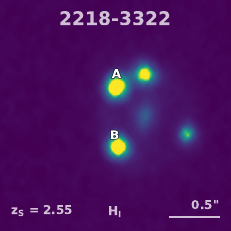}
\includegraphics[width=0.16255\textwidth]{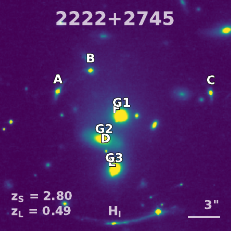}
\includegraphics[width=0.16255\textwidth]{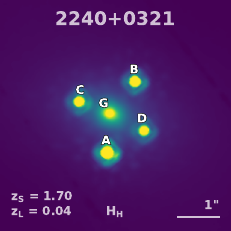}
\includegraphics[width=0.16255\textwidth]{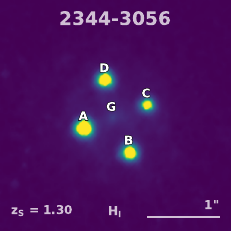}
\end{figure*}

\end{appendix}
\end{document}